\documentclass[a4paper]{article}
\usepackage{eurosym}
\usepackage{graphicx}
\usepackage{amsmath}
\usepackage{amsfonts}
\usepackage{amssymb}
\usepackage[caption=false]{subfig}
\usepackage{float}

\begin{document}

\title{Multi-soliton states under triangular spatial modulation of the
quadratic nonlinearity}
\author{Vitaly Lutsky$^{1}$ and Boris A. Malomed$^{1,2}$ \\
$^{1}$Department of Physical Electronics, School of Electrical Engineering,\\
Tel Aviv University, Tel Aviv 69978, Israel\\
$^{2}$ITMO University, St. Petersburg 197101, Russia}
\date{\today{}}
\maketitle

\begin{abstract}
{\large \ }We introduce multi-soliton sets in the two-dimensional medium
with the $\chi ^{(2)}$ nonlinearity subject to spatial modulation in the
form of a triangle of singular peaks. Various families of symmetric and
asymmetric sets are constructed, and their stability is investigated. Stable
symmetric patterns may be built of $1$, $4$, or $7$ individual solitons,
while stable asymmetric ones contain $1$, $2$, or $3$ solitons. Symmetric
and asymmetric patterns may demonstrate mutual bistability. The shift of the
asymmetric single-soliton state from the central position is accurately
predicted analytically. Vortex rings composed of three solitons are produced
too.
\end{abstract}

\pagenumbering{arabic}

\section{Introduction}

The significance of solitons (self-trapped modes) in various areas of
physics is commonly known \cite{Agrawal,Peyrard}. Recently, considerable
attention has been drawn to possibilities to extend the variety of solitons
in nonlinear media by imposing spatial modulation on the local strength of
the nonlinear interaction, i.e., the creation of effective nonlinear
potentials for solitons (alias \textit{pseudopotentials}, as they were
originally called in solid-state physics \cite{pseudo}). Many results
produced in this direction were collected in review \cite{review}. A
majority of studies performed in this area dealt with the ubiquitous cubic,
alias $\chi ^{(3)}$, nonlinearity, which finds commonly known realizations
in optics, in the form of self-focusing of optical beams due to the Kerr
effect \cite{Agrawal}, and as collisional interactions in atomic
Bose-Einstein condensates (BECs) \cite{BEC}. In optical media, spatial
modulation of the local Kerr coefficient can be induced by means of
inhomogeneous distributions of resonant dopants, which enhance the $\chi
^{(3)}$ nonlinearity via two-photon resonance \cite{Kip}. In BEC, a similar
effect can be produced by means of inhomogeneous optical \cite%
{BEC-optical1,BEC-optical2} or magnetic \cite{BEC-magnetic} fields, as well
as a combination of both \cite{BEC-combined}, which affect the local
strengths of the collisional nonlinearity via the Feshbach resonance.

A specific ramification of such models in one dimension (1D) is based on a
locally singular modulation of the self-focusing in the form of a cusp, with
coefficient $\chi ^{(3)}\sim |x|^{-\alpha }$, which was introduced in Ref.
\cite{Olga}. The relevant range of values of the singular-modulation power
is $0\leq \alpha <1$. This model offers a possibility to generalize the
study of the onset of collapse in nonlinear media \cite{Berge,Fibich}, as
well as to emulate the nonlinear dynamics in a sub-1D\ space, with the
effective dimension $D=2\left( 1-\alpha \right) /\left( 2-\alpha \right) <1$
\cite{Olga}. The singular modulation can be emulated by means of
above-mentioned techniques, tuning them to the exact resonance in a narrow
layer.

Another possibility is to consider spatial modulation of the local
interaction strength in media with the quadratic ($\chi ^{(2)}$)
nonlinearity, which has well-known realizations in optics \cite{Hagan}-\cite%
{Ady}. Experimental realizations of such settings are possible, in
particular, using the well-elaborated technique of the quasi-phase-matching
\cite{QPM1}-\cite{QPM3}, which can be implemented in a spatially nonuniform
form, in 1D and 2D geometries alike, thus helping one to create a required
profile of the $\chi ^{(2)}$ coefficient \cite{QPM-landscape}-\cite%
{QPM-landscape3}. In the theoretical analysis, singular modulation of the
quadratic nonlinearity in the 1D system, accounted for by a delta-function, $%
\chi ^{(2)}(x)\sim \delta (x)$, and localized modes (solitons) pinned to it,
were introduced in Ref. \cite{Canberra}, and a pair of modulating
delta-functions was considered in Ref. \cite{Asia}. A discrete version of
the localized quadratic\ nonlinearity was elaborated in the form of a linear
lattice with one or two $\chi ^{(2)}$-nonlinear sites embedded in it \cite%
{Valery-chi2}.

While the delta-like modulation of the local $\chi ^{(2)}$ coefficient may
not be easily realized in the experiment, a more realistic case of the 1D
singular modulation, with $\chi ^{(2)}\sim |x|^{-\alpha }$ and positive $%
\alpha $, was introduced in Ref. \cite{we-quadratic}. It was found that this
modulation format supports quadratic solitons, pinned to the singular peak,
for $\alpha <1$ (the pinned modes vanish at $\alpha =1$), and they are
chiefly stable. A natural extension of that setting is a symmetric pair of
two peaks (similar to the above-mentioned symmetric set of two
delta-functions multiplying the nonlinear terms \cite{Asia}). The
consideration of the twin peaks has made it possible to predict effects of
the spontaneously symmetry breaking \cite{breaking} and formation of
asymmetric two-soliton states pinned to the two peaks \cite{we-quadratic}.
These results may be used for the design of steering optical beams in the
form of spatial solitons.

An essential advantage of using the quadratic nonlinearity with this type of
the local modulation is that it may be extended to the 2D geometry, by
choosing $\chi ^{(2)}(r)\sim r^{-\alpha }$ ($r$ is the radial coordinate),
while any 2D singular modulation of the self-focusing $\chi ^{(3)}$
nonlinearity leads to collapse. In terms of optics, the 2D system models
light propagation in bulk media with the quadratic nonlinearity. In
particular, the model admits stable spatial solitons, which represent
self-trapped light beams in the bulk medium \cite{Hagan}-\cite%
{Quadratic_solitons}. The analysis of the 2D setting, performed in Ref. \cite%
{we-quadratic}, has revealed that the solitons, pinned to the
singular-modulation peak, exist at $\alpha <2$, and they have a stability
region at $\alpha <0.5$. Solitons with embedded vorticity can also be pinned
to the singularity peak, but they all are unstable.

The availability of stable 2D solitons attached to the singular-modulation
peak suggests a possibility to study multi-soliton patterns in multi-peak
modulation profiles, subject to natural symmetry conditions. The simplest
profile which realizes the 2D symmetry is an equilateral set of three peaks,
cf. Refs. \cite{trimer1}-\cite{trimer6}. This is the subject of the present
paper, which is structured as follows. The model is introduced in Section
II. In Section III, we report numerical results for\ the existence and
stability of various patterns, which may be classified as symmetric and
asymmetric ones, with respect to the underlying triangular modulation
structure. Some results are also obtained in an analytical form, \textit{viz}%
., prediction of an asymmetric location of a single soliton between the
three singular cusps. Section IV deals with ring-shaped three-soliton sets
which carry the phase circulation of $2\pi $, i.e., composite vortices
admitted by the three-peak modulation structure, cf. Refs. \cite{vort1}-\cite%
{vort4}. The paper is concluded by Section V.

\section{The model}

Dynamical models for self-guided beams in $\chi ^{(2)}$ media have been
studied in detail, as summarized in reviews \cite{Hagan}-\cite{Ady}. In the
present work, we focus on the basic case of the two-wave (degenerate, alias
Type-I) quadratic interactions in the bulk medium, which are modeled by the
system of 2D spatial-domain equations for complex amplitudes of the
fundamental-frequency (FF) and second-harmonic (SH) waves, $u\left(
x,y,z\right) $ and $v(x,y,z)$. Here, following the above-mentioned direction
of studies of the nonlinear wave propagation in media with a locally
modulated nonlinearity strength \cite{review}, we consider the system with
an $\left( x,y\right) $-dependent $\chi ^{(2)}$ coefficient. As also said
above, the case of singular modulation is an especially interesting one.
Thus, the 2D system with the cusp-shaped modulation of the $\chi ^{(2)}$
coefficient is written, in the scaled form, as

\begin{gather}
iu_{z}+\frac{1}{2}\nabla ^{2}u+{r}^{-\alpha }u^{\ast }v=0,  \label{u_2d} \\
2iv_{z}+\frac{1}{2}\nabla ^{2}v-Qv+\frac{1}{2}{r}^{-\alpha }u^{2}=0,
\label{v_2d}
\end{gather}%
where $z$ is the propagation distance, diffraction operator $(1/2)\nabla
^{2} $ acts on transverse coordinates $\left( x,y\right) $, $r\equiv \sqrt{%
x^{2}+y^{2}}$, the asterisk stands for the complex conjugate, and real
coefficient $Q$ represents the SH-FF mismatch. By means of obvious
rescaling, we fix the mismatch parameter to take one of three values,%
\begin{equation}
Q=0,+1,-1.  \label{Q}
\end{equation}%
In fact, similar results are obtained for all the three values defined in
Eq. (\ref{Q}).

Localized solutions of Eqs. (\ref{u_2d}) and (\ref{v_2d}) are characterized
by the total power (alias the Manley-Rowe invariant),%
\begin{equation}
P=\int \int \left[ |u\left( x,y\right) |^{2}+4|v\left( x,y\right) |^{2}%
\right] dxdy,  \label{P}
\end{equation}%
which is a dynamical invariant of the system, along with its Hamiltonian,%
\begin{equation}
H=\int \int \left\{ \frac{1}{2}\left( \left\vert \nabla u\right\vert
^{2}+\left\vert \nabla v\right\vert ^{2}\right) -\frac{1}{2}r^{-\alpha }%
\left[ u^{2}v^{\ast }+(u^{\ast })^{2}v\right] +Q|v|^{2}\right\} dxdy.
\label{H}
\end{equation}%
The isotropic configuration with the single-peak modulation conserves the
angular momentum too. However, the triangular setting considered below is
anisotropic, breaking the angular-momentum conservation.

Positive exponent $\alpha $ in Eqs. (\ref{u_2d}) and (\ref{v_2d}) determines
the form of the singularity, $\alpha =0$ corresponding to the uniform
medium. As mentioned above, a basic result reported in Ref. \cite%
{we-quadratic} is that these equations generate fundamental (zero-vorticity)
localized states pinned to the singular-modulation peak at $\alpha <2$, and
these states have a stability area at $\alpha <0.5$. Under this condition,
they tend to be stable and unstable, respectively, at smaller and larger
values of the total power, $P$, at all values of the mismatch, see Eq. (\ref%
{Q}).

Stationary solutions to Eqs. (\ref{u_2d}), (\ref{v_2d}) with real
propagation constant $k$ are looked for as
\begin{equation}
\left\{ u(x,y,z),v(x,y,z)\right\} =\left\{ e^{ikz}\varphi (x,y),e^{2ikz}\psi
(x,y)\right\} ,  \label{stationary}
\end{equation}%
with functions $\varphi (x,y)$ and $\psi (x,y)$ (which are complex for
vortex modes) obeying the stationary equations:

\begin{gather}
-k\varphi +\frac{1}{2}\nabla ^{2}\varphi +{r}^{-\alpha }\varphi ^{\ast }\psi
=0,  \label{phi_2d_stationary} \\
-4k\psi +\frac{1}{2}\nabla ^{2}\psi -Q\psi +\frac{1}{2}{r}^{-\alpha }\varphi
^{2}=0.  \label{psi_2d_stationary}
\end{gather}%
The dependence between the total power and $k$, obtained in a numerical form
from Eqs. (\ref{phi_2d_stationary}) and (\ref{psi_2d_stationary}), satisfies
the Vakhitov-Kolokolov criterion, $dP/dk>0$ (for all values of mismatch, $Q$%
), which is a well-known necessary, but not sufficient, stability criterion
for solitons \cite{VK,Berge,Fibich}.

Solutions to Eqs. (\ref{phi_2d_stationary}) and (\ref{psi_2d_stationary})
may also be looked for as vortices with integer topological charge $m$ in
its FF components (and charge $2m$ in the SH component) \cite{Dima11,Dima21}%
, i.e.,%
\begin{equation}
\varphi =e^{im\theta }U(r),\psi =e^{2im\theta }V(r),  \label{UV}
\end{equation}%
in terms of the polar coordinates ($r,\theta $), with real radial functions $%
U$ and $V$ obeying the following equations:

\begin{gather}
-kU+\frac{1}{2}\left( \frac{d^{2}}{dr^{2}}+\frac{1}{r}\frac{d}{dr}-\frac{%
m^{2}}{r^{2}}\right) U+{r}^{-\alpha }VU=0,  \label{Ur} \\
-4kV+\frac{1}{2}\left( \frac{d^{2}}{dr^{2}}+\frac{1}{r}\frac{d}{dr}-\frac{%
m^{2}}{r^{2}}\right) V-QV+\frac{1}{2}{r}^{-\alpha }U^{2}=0,  \label{Vr}
\end{gather}%
supplemented by boundary conditions $U\sim r^{|m|}$ and $V\sim r^{2|m|}$ at $%
r\rightarrow 0$. Unlike the fundamental modes, all the vortical ones
supported by the single peak were found to be unstable \cite{we-quadratic},
similar to the well-known instability of 2D solitons with embedded vorticity
in the uniform $\chi ^{(2)}$ medium \cite{Dima11}-\cite{Dima12}.

The subject of the consideration in the present work is the set of three
equally separated singular-modulation peaks, which is represented by the
following \textquotedblleft triangular" modulation profile, labeled by
symbol $\Delta $:

\begin{gather}
\chi _{\triangle }^{(2)}(\alpha ,X_{A,B,C},Y_{A,B,C})=\left[
(x-X_{A})^{2}+(y-Y_{A})^{2}\right] ^{-\frac{\alpha }{2}}+  \notag \\
\left[ (x-X_{B})^{2}+(y-Y_{B})^{2}\right] ^{-\frac{\alpha }{2}}+\left[
(x-X_{C})^{2}+(y-Y_{C})^{2}\right] ^{-\frac{\alpha }{2}}.  \label{R}
\end{gather}%
Here $\left( X_{A,B,C},Y_{A,B,C}\right) $ are coordinates of the three
peaks, each located at distance $R$ from the center, as shown in Fig.(\ref%
{fig1}). The respective system of the coupled FF-SH propagation equations is
written as
\begin{gather}
i{u}_{z}+\frac{1}{2}\nabla ^{2}u+\chi _{\triangle }^{(2)}(\alpha
,X_{A,B,C},Y_{A,B,C})u^{\ast }v=0,  \label{u_2d_triangle} \\
i{v}_{z}+\frac{1}{2}\nabla ^{2}v-Qv+\frac{1}{2}\chi _{\triangle
}^{(2)}(\alpha ,X_{A,B,C},Y_{A,B,C})u^{2}=0.  \label{v_2d_triangle}
\end{gather}

\begin{figure}[tbp]
\centering
\captionsetup[subfigure]{labelformat=empty}
\subfloat[]{
		\includegraphics[width=0.7\textwidth]{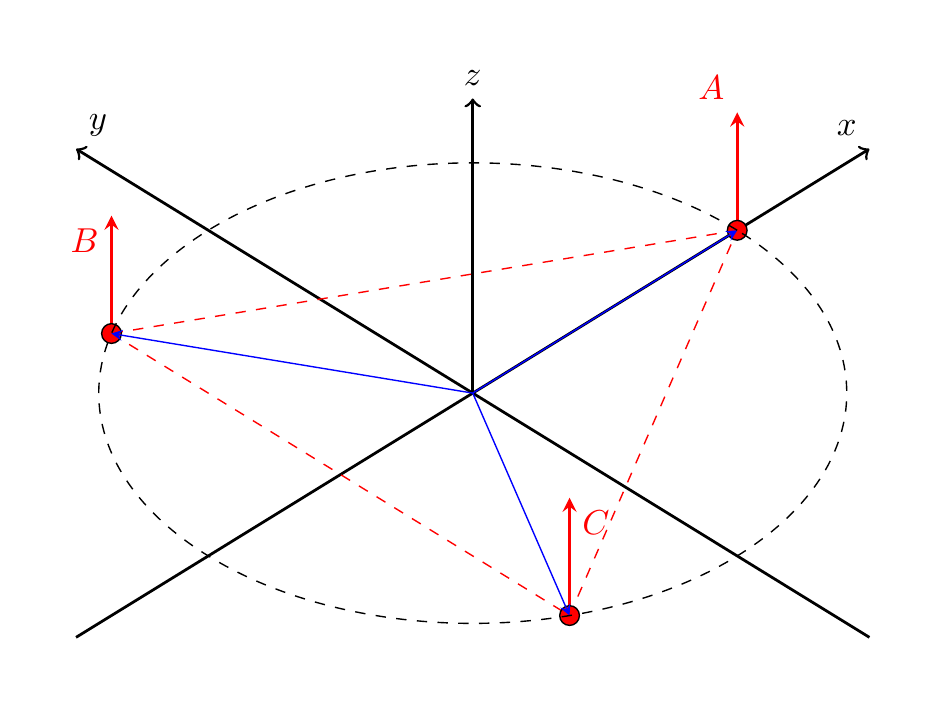}
	}
\caption{The set of three singularity peaks A, B, and C of the nonlinearity
modulation, each one separated by distance $R$ from the center.}
\label{fig1}
\end{figure}

The stationary version of Eqs. (\ref{u_2d_triangle}) and \ref{v_2d_triangle}%
) is

\begin{gather}
-k\varphi +\frac{1}{2}\nabla ^{2}\varphi +\chi _{\triangle }^{(2)}(\alpha
,X_{A,B,C},Y_{A,B,C})\varphi ^{\ast }\psi =0,
\label{phi_2d_stationary_triangle} \\
-4k\psi +\frac{1}{2}\nabla ^{2}\psi -Q\psi +\frac{1}{2}\chi _{\triangle
}^{(2)}(\alpha ,X_{A,B,C},Y_{A,B,C})\varphi ^{2}=0,
\label{psi_2d_stationary_triangle}
\end{gather}%
cf. Eqs. (\ref{phi_2d_stationary}) and (\ref{psi_2d_stationary}). Solutions
to Eqs. (\ref{phi_2d_stationary_triangle}) and (\ref%
{psi_2d_stationary_triangle}) were produced by means of two numerical
schemes, \textit{viz}., the Newton's conjugate gradient method, and the
squared operator method \cite{Yang}, which are appropriate techniques in
this context. In particular, the Newton's algorithm provides convergence of
the numerical solutions after fewer iterations than demanded by other
numerical schemes.

The stability analysis for stationary states was performed by taking
perturbed solutions in the usual form \cite{Yang,Pelinovsky},
\begin{eqnarray}
u &=&\left[ \varphi (x,y)+\epsilon _{u}(x,y,z)\right] e^{ikz},
\label{u_Perturbed_2d} \\
v &=&\left[ \psi (x,y)+\epsilon _{v}(x,y,z)\right] e^{2ikz},
\label{v_Perturbed_2d}
\end{eqnarray}%
with perturbation eigenmodes looked for as
\begin{equation}
\begin{cases}
\epsilon _{u}=\xi _{u}^{+}(x,y)e^{\lambda z}+\xi _{u}^{\ast
-}(x,y)e^{-\lambda ^{\ast }z}, \\
\epsilon _{v}=\xi _{v}^{+}(x,y)e^{\lambda z}+\xi _{v}^{\ast
-}(x,y)e^{-\lambda ^{\ast }z},%
\end{cases}
\label{J}
\end{equation}%
where $\lambda $ is the respective eigenvalue (that may be complex),
instability corresponding to $\mathrm{Re}(\lambda )\neq 0$. The
corresponding eigenvalue problem for the linearized equations was then
solved by means of the spectral collocation method \cite{Yang}. Finally, the
so predicted stability was checked in direct simulations of the evolution of
perturbed solutions of Eqs. (\ref{phi_2d_stationary_triangle}) and (\ref%
{psi_2d_stationary_triangle}). The simulations were run by means of the
fourth-order Runge-Kutta algorithm, implemented in the 2D Cartesian
coordinates.

\section{Numerical results: Zero-vorticity patterns}

\subsection{The fundamental mode pinned to the single peak}

Zero-vorticity (alias fundamental) states were found as numerical solutions
of Eqs. (\ref{phi_2d_stationary_triangle}) and (\ref%
{psi_2d_stationary_triangle}), with the help of the above-mentioned Newton's
and conjugate-gradient method, following the pattern presented in book \cite%
{Yang}. Before proceeding to the consideration of the triangular scheme, it
was necessary to create appropriate building blocks, in the form of
solutions pinned to the single modulation peak, i.e., solutions in the form
of expressions (\ref{UV}), with the radial functions satisfying Eqs. (\ref%
{Ur}) and (\ref{Vr}). The latter equations were solved starting from the
input in the form of

\begin{equation}
U(r)=Ar^{m}\exp \left( -\rho r^{2}\right) ,V(r)=Br^{2m}\exp \left( -\gamma
r^{2}\right) ,  \label{2d_ansatz}
\end{equation}%
with constants $A$, $B$ and $\rho >0$, $\gamma >0$.

Note that the same expressions (\ref{2d_ansatz}) can be used as the ansatz
for constructing the VA (variational approximation), using the Lagrangian of
Eqs. (\ref{Ur}) and (\ref{Vr}),

\begin{gather}
L=\int_{0}^{\infty }rdr\left\{ \frac{1}{2}\left[ \left( \frac{dU}{dr}\right)
^{2}+\left( \frac{dV}{dr}\right) ^{2}\right] \right.  \notag \\
\left. +\left[ \left( k+\frac{m^{2}}{2r^{2}}\right) U^{2}+\left( 4k+\frac{%
2m^{2}}{r^{2}}+Q\right) V^{2}-r^{-\alpha }U^{2}V\right] \right\} .
\label{L2D}
\end{gather}%
The substitution of \textit{ansatz} (\ref{2d_ansatz}) in the Lagrangian
yields the following effective Lagrangian:

\begin{equation}
\begin{split}
L& =\frac{1}{2}\left\{ 4^{-m}B^{2}m(4k+Q)\gamma ^{-1-2m}\Gamma \left(
2m\right) \right. \\
& \left. +\frac{1}{2}4^{-m}B^{2}\gamma ^{-2m}\Gamma \left( 2+2m\right)
\right\} + \\
& \frac{1}{4}A^{2}\left\{ 2^{-m}\rho ^{-1-m}\left[ k\Gamma \left( 1+m\right)
+\rho \Gamma \left( 2+m\right) \right] \right. \\
& \left. -2B(\gamma +2\rho )^{-1-2m+\frac{\alpha }{2}}\Gamma \left( 1+2m-%
\frac{\alpha }{2}\right) \right\} ,
\end{split}
\label{Lagrangian_2d}
\end{equation}%
where $\Gamma $ is the Euler's Gamma-function. In the framework of the VA,
amplitudes $A$, $B$ and inverse square widths of the two components, $\rho $
and $\gamma $, were found, for given $k$ and $m$, from the Euler-Lagrange
equations, $\partial L/\partial B=\partial L/\partial \left( A^{2}\right)
=\partial L/\partial \rho =\partial L/\partial \gamma =0$, applied to
Lagrangian (\ref{Lagrangian_2d}). This system of algebraic equations was
solved numerically.

The practical significance of the VA is in the fact that it predicts the
mode pinned to an individual peak in the explicit analytical form (\ref%
{2d_ansatz}). Then, appropriate superpositions of such modes, with separated
centers, can be immediately used as inputs for generating multi-mode
patterns pinned to the set of three peaks. This approach works efficiently
even in cases when the accuracy of the VA-predicted mode for the single peak
is relatively poor, in comparison with the corresponding full numerical
solution.

\subsection{Symmetric fundamental modes in the three-peak configuration}

The numerical solution of Eqs. (\ref{phi_2d_stationary_triangle}) and (\ref%
{psi_2d_stationary_triangle}) reveals many species of stationary patterns
pinned to the three-peak modulation profiles, both stable and unstable ones.
Some stationary states share the three-fold symmetry of the underlying
triangular profile. They were found to be stable in three cases:

\begin{itemize}
\item Single-soliton states, placed at the midpoint (central position)
between the three peaks. Their shape is very similar to the well-known
stable 2D fundamental solitons in uniform $\chi ^{(2)}$ media \cite%
{Quadratic_solitons_Malomed,Quadratic_solitons}, see an example in Fig \ref%
{fig:Stable_1pulse_symmetric}.

\item Four-soliton sets, composed of three identical solitons placed close
to the three modulation peaks, and an additional soliton, with the sign of
its FF component opposite to that of the other three constituents of the
set, which is placed at the midpoint (the sign of the SH component is the
same for all the four solitons, and their local peak powers, i.e., maximum
values of $|\varphi \left( x,y\right) |^{2}$ and $|\psi \left( x,y\right)
|^{2}$, are nearly equal too). The opposite signs of the FF components make
the interaction between the central soliton and ones belonging to the
surrounding triad repulsive \cite{potential}, which is compensated by
attraction of the latter solitons to neighboring singularities of
nonlinearity strength. An example of a stable four-soliton patterns in
displayed in Fig. \ref{fig:Stable_4pulses_symmetric}.

\item Seven-soliton sets, composed, as above, of three identical solitons
placed very close to the three modulation peaks, an additional soliton,
placed at the midpoint, and three other ones, which are located on the
extension of lines connecting the midpoint with each peak, see an example in
Fig. \ref{fig:Stable_7pulses_symmetric}. In this case too, the sign of the
FF component of the four additional solitons is opposite to that of the
triad directly pinned to the singular-modulation peaks, while the signs of
the SH\ component are the same for all the seven constituents, and their
peak powers are nearly equal.
\end{itemize}

\begin{figure}[tbp]
\centering
\subfloat[]{
		\includegraphics[width=0.45\textwidth]{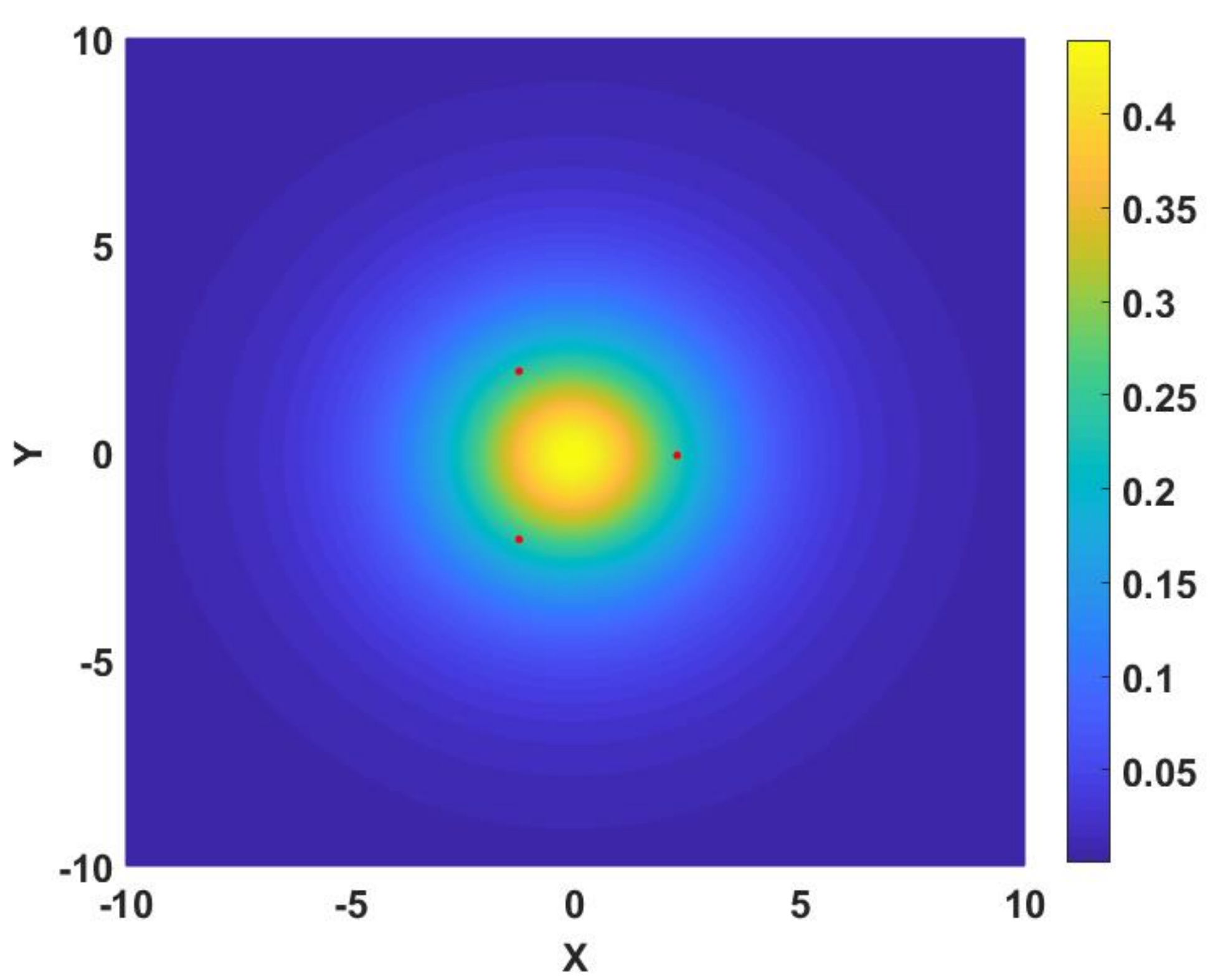}
	} \subfloat[]{
		\includegraphics[width=0.45\textwidth]{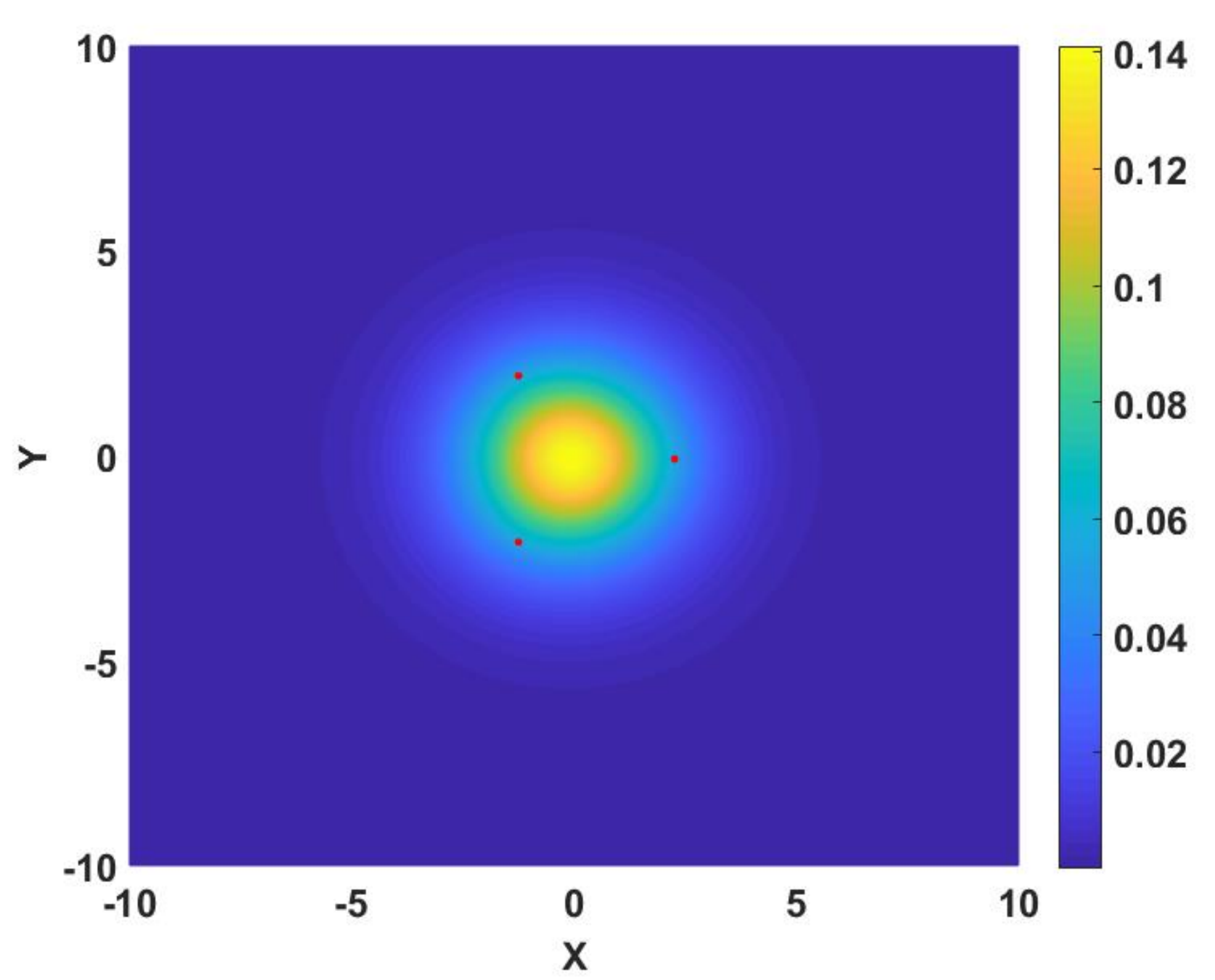}
	}
\caption{(Color online) A stable symmetric single-soliton pattern, with the
following values of parameters in Eqs. (\protect\ref{R})-(\protect\ref%
{psi_2d_stationary_triangle}): the modulation-singularity power $\protect%
\alpha =0.15$, mismatch $Q=+1$, and distance of each singularity peak from
the center $R=2.3438$ (here and in similar plots displayed below, red dots
show location of the singularities). The propagation constant corresponding
to this soliton set is $k=0.1$. Panels (a) and (b) display the stationary FH
and SH fields, $\protect\varphi \left( x,y\right) $ and $\protect\psi \left(
x,y\right) $, respectively.}
\label{fig:Stable_1pulse_symmetric}
\end{figure}

\begin{figure}[tbp]
\centering
\subfloat[]{
		\includegraphics[width=0.45\textwidth]{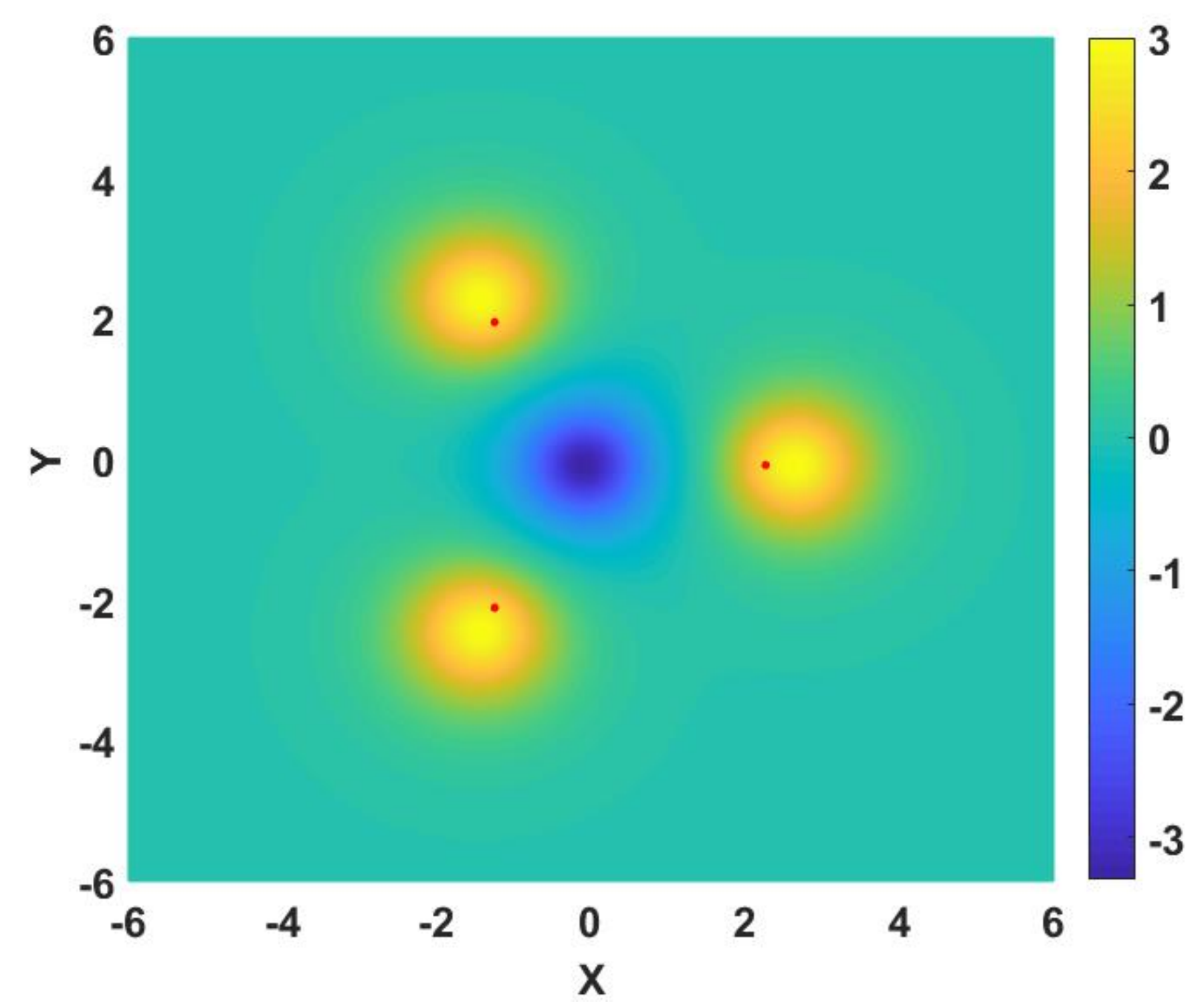}
	} \subfloat[]{
		\includegraphics[width=0.45\textwidth]{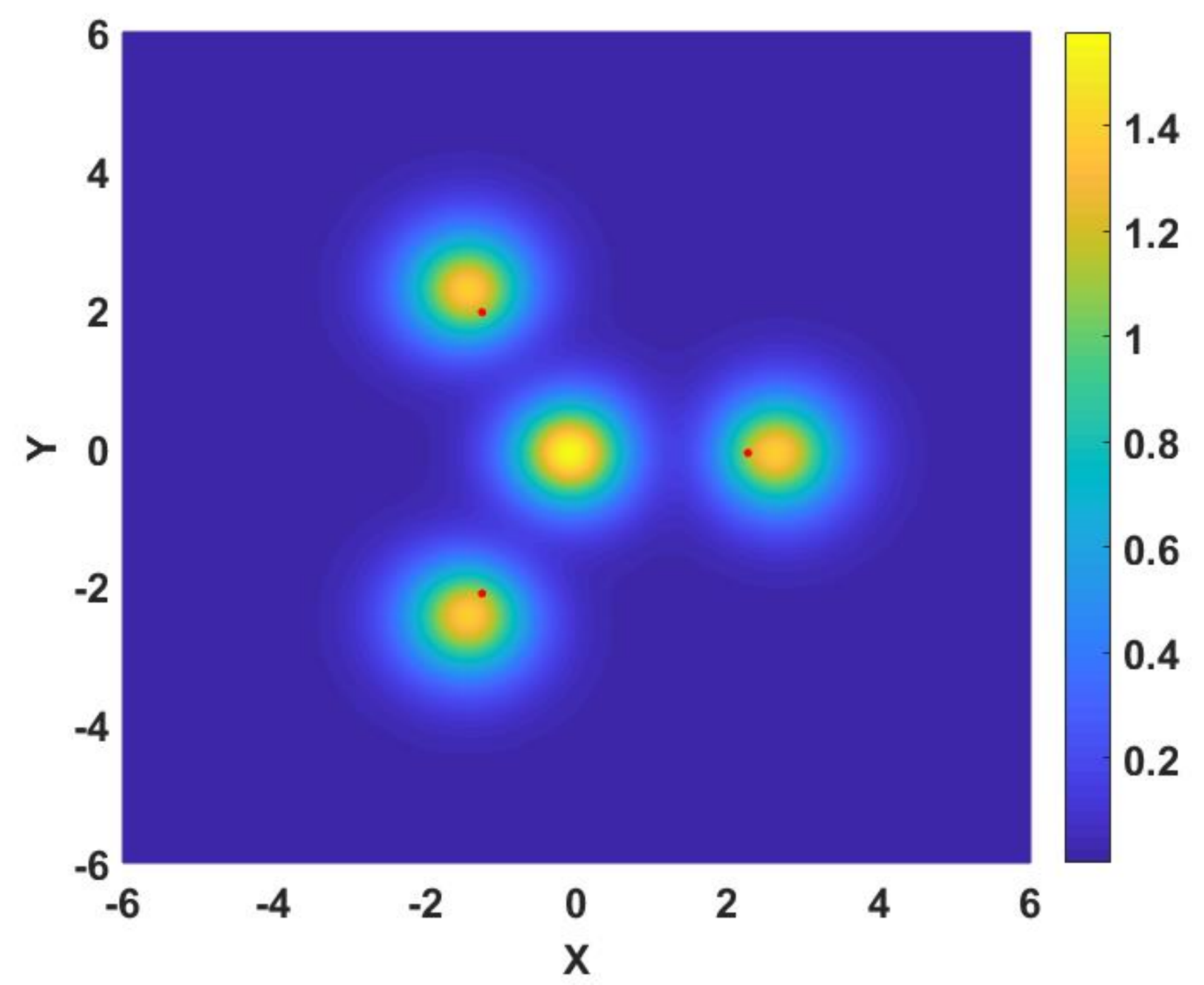}
	}
\caption{(Color online) A stable set of four\ solitons, with parameters $%
\protect\alpha =0.15$, $Q=+1$, and $R=2.3438$. The propagation constant
corresponding to this soliton set is $k=1.1$. Panels (a) and (b) display the
FH and SH fields, $\protect\varphi \left( x,y\right) $ and $\protect\psi %
\left( x,y\right) $, respectively.}
\label{fig:Stable_4pulses_symmetric}
\end{figure}

\begin{figure}[tbp]
\centering
\subfloat[]{
		\includegraphics[width=0.45\textwidth]{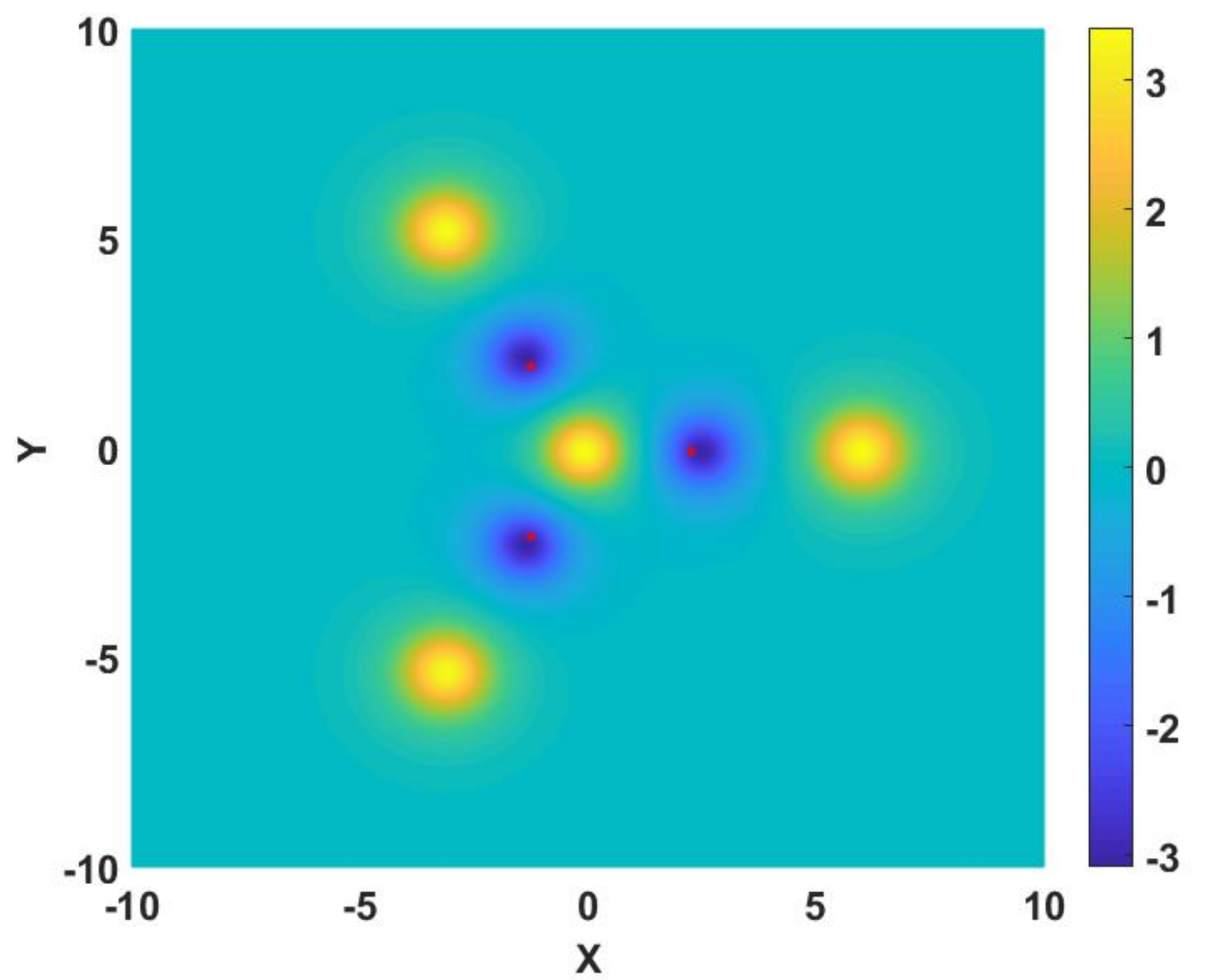}
	} \subfloat[]{
		\includegraphics[width=0.45\textwidth]{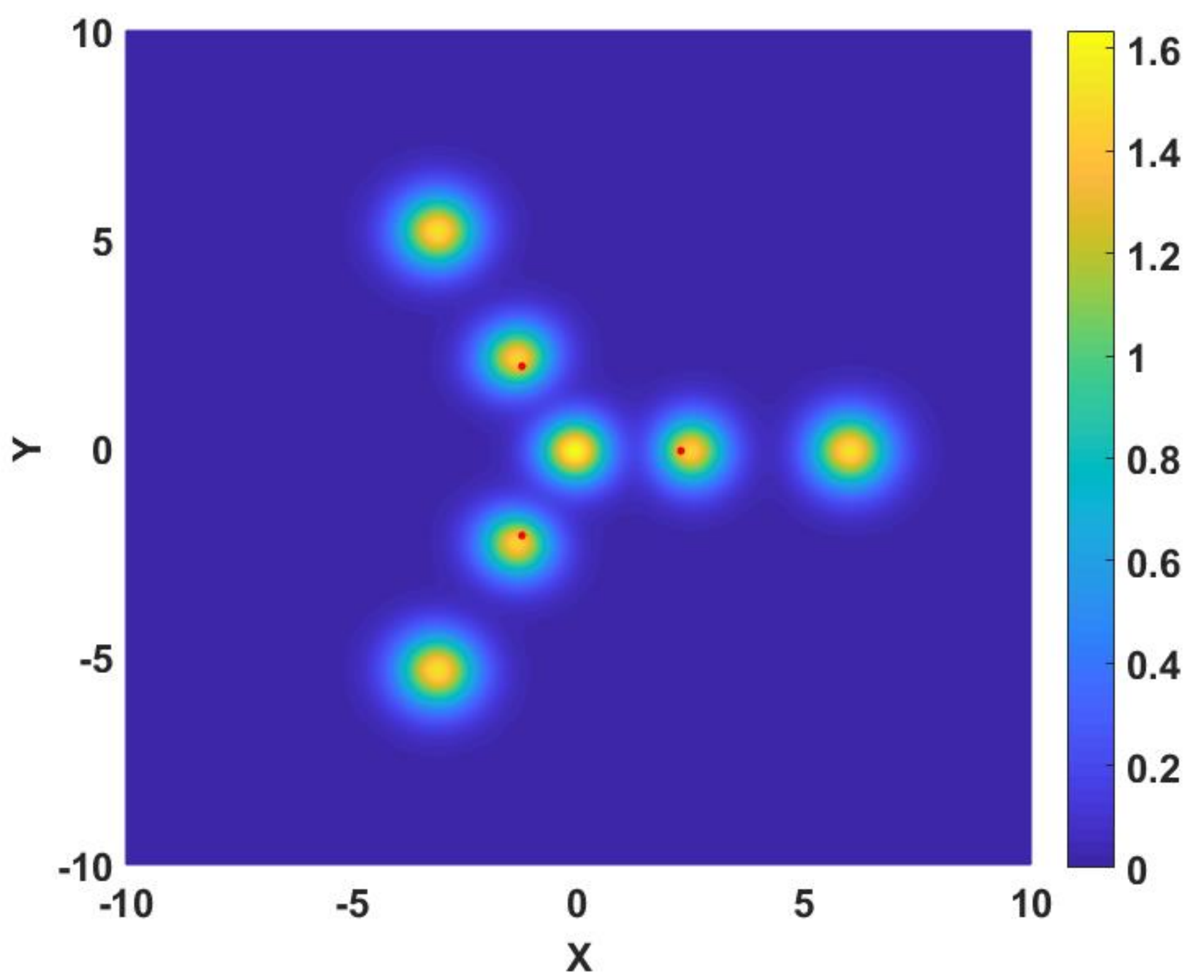}
	}
\caption{(Color online) The same as in Fig. \protect\ref%
{fig:Stable_4pulses_symmetric}, but for a stable set of seven solitons, with
parameters $\protect\alpha =0.15,Q=1,R=2.3438$, and $k=1.1$.}
\label{fig:Stable_7pulses_symmetric}
\end{figure}

\begin{figure}[tbp]
\centering
\subfloat[]{
		\includegraphics[width=0.45\textwidth]{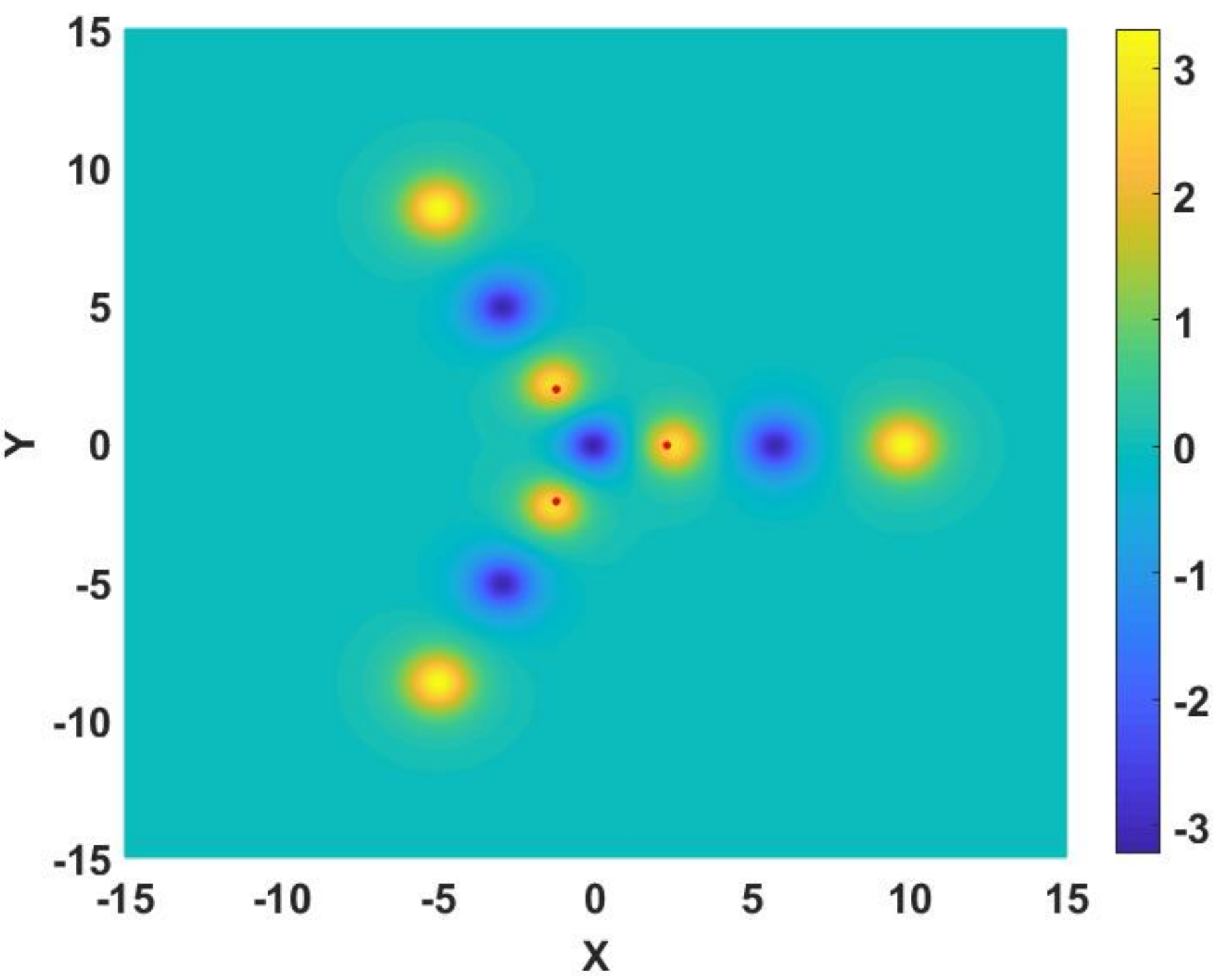}
	} \subfloat[]{
		\includegraphics[width=0.45\textwidth]{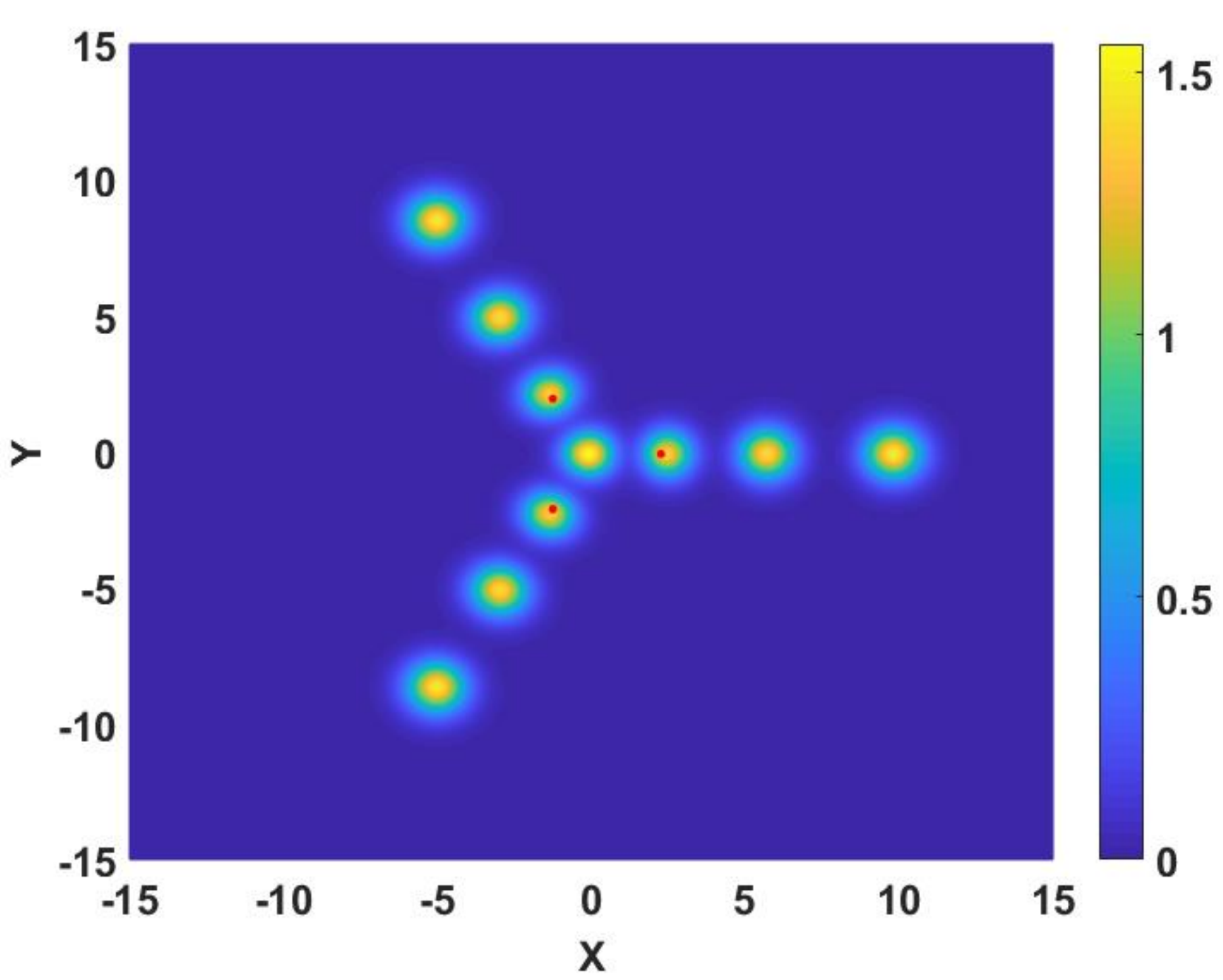}
	}
\caption{(Color online) The same as in Figs. \protect\ref%
{fig:Stable_4pulses_symmetric} and \protect\ref{fig:Stable_7pulses_symmetric}%
, but for an unstable symmetric set of ten solitons, with parameters $%
\protect\alpha =0.15$, $Q=1$, $k=1$, and $R=2.3438$.}
\label{fig:Unstable_10pulses_symmetric}
\end{figure}

\begin{figure}[tbp]
\centering
\subfloat[]{
		\includegraphics[width=0.49\textwidth]{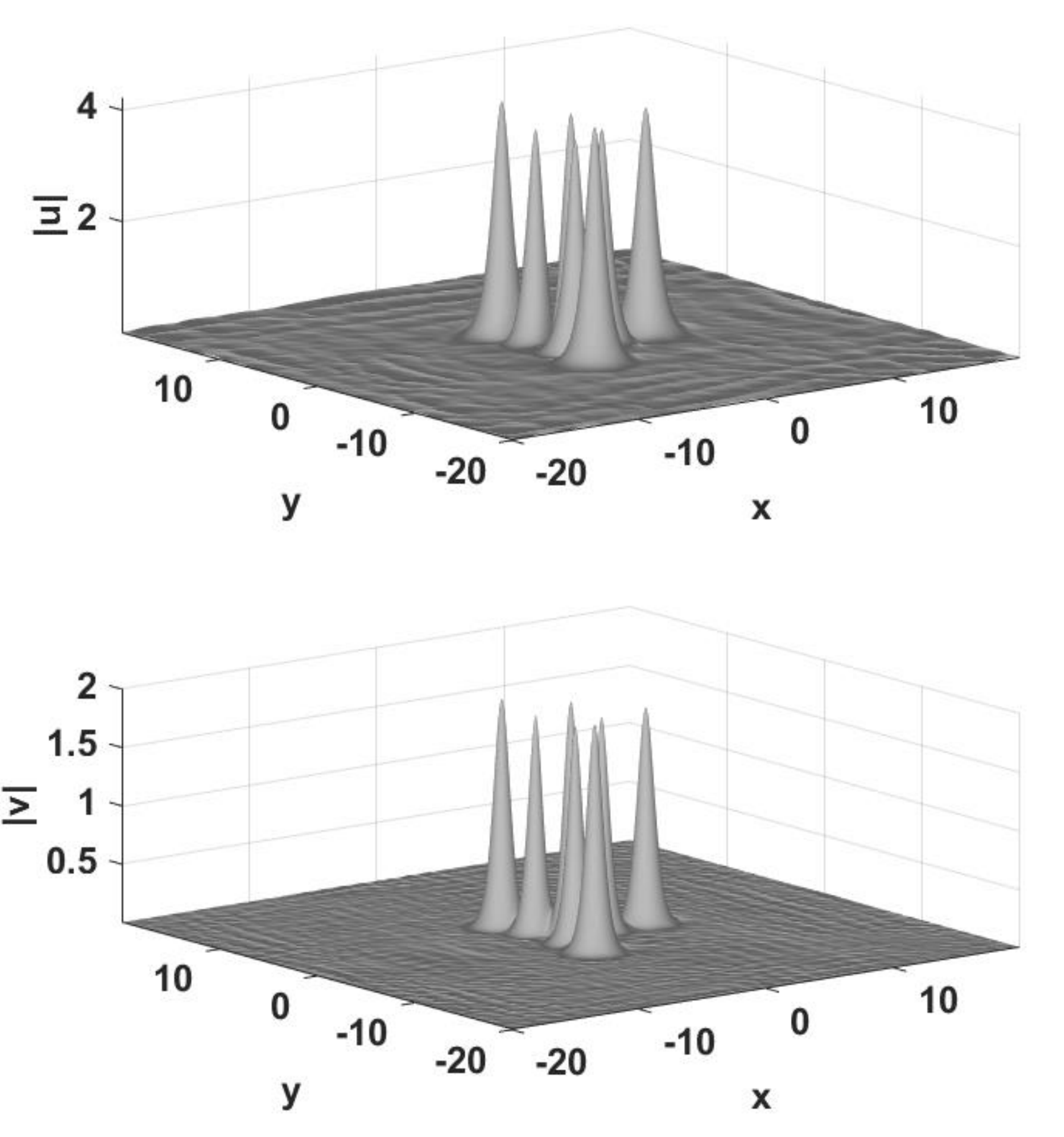}
	} \subfloat[]{
		\includegraphics[width=0.49\textwidth]{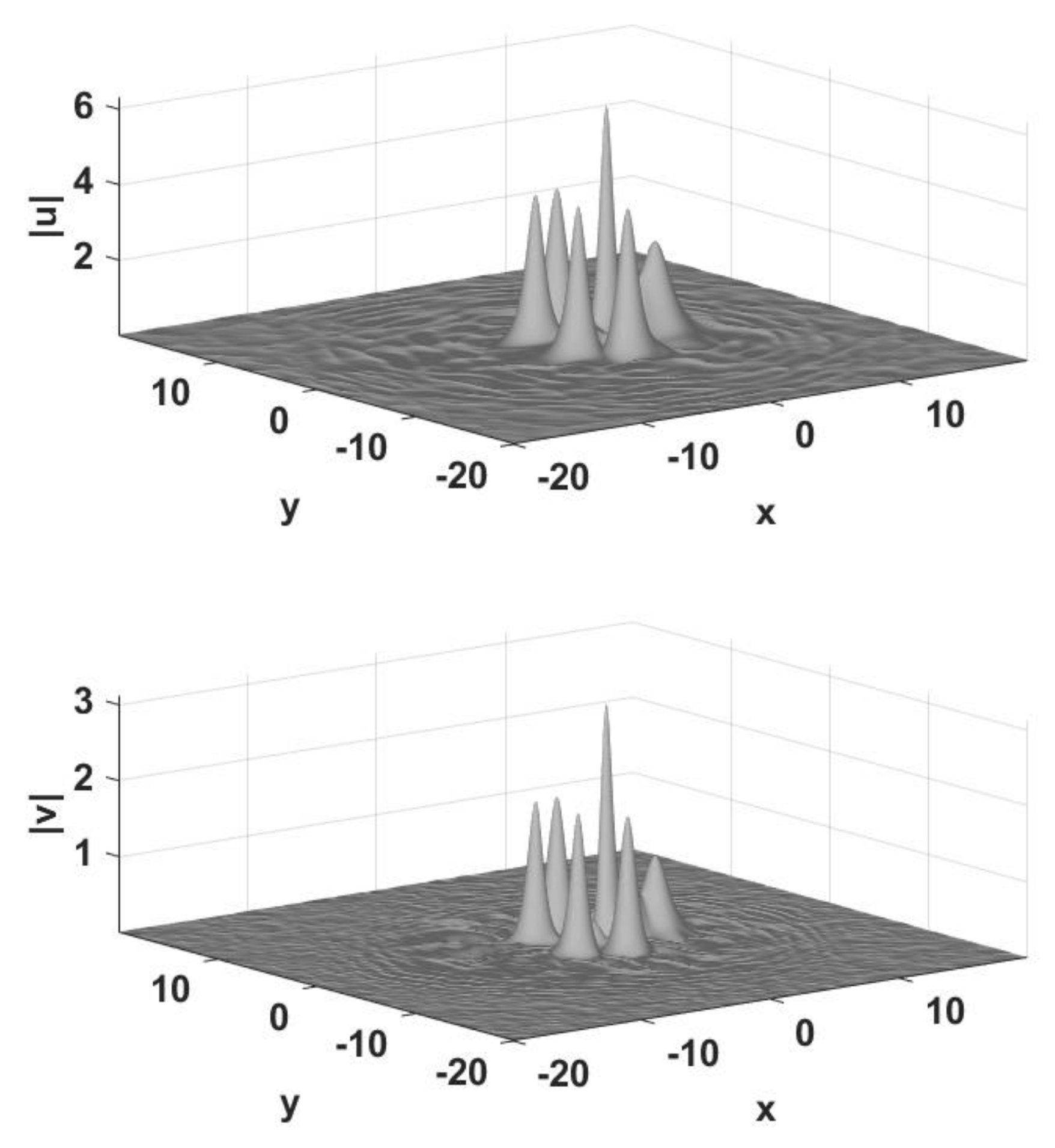}
	} \newline
\subfloat[]{
		\includegraphics[width=0.49\textwidth]{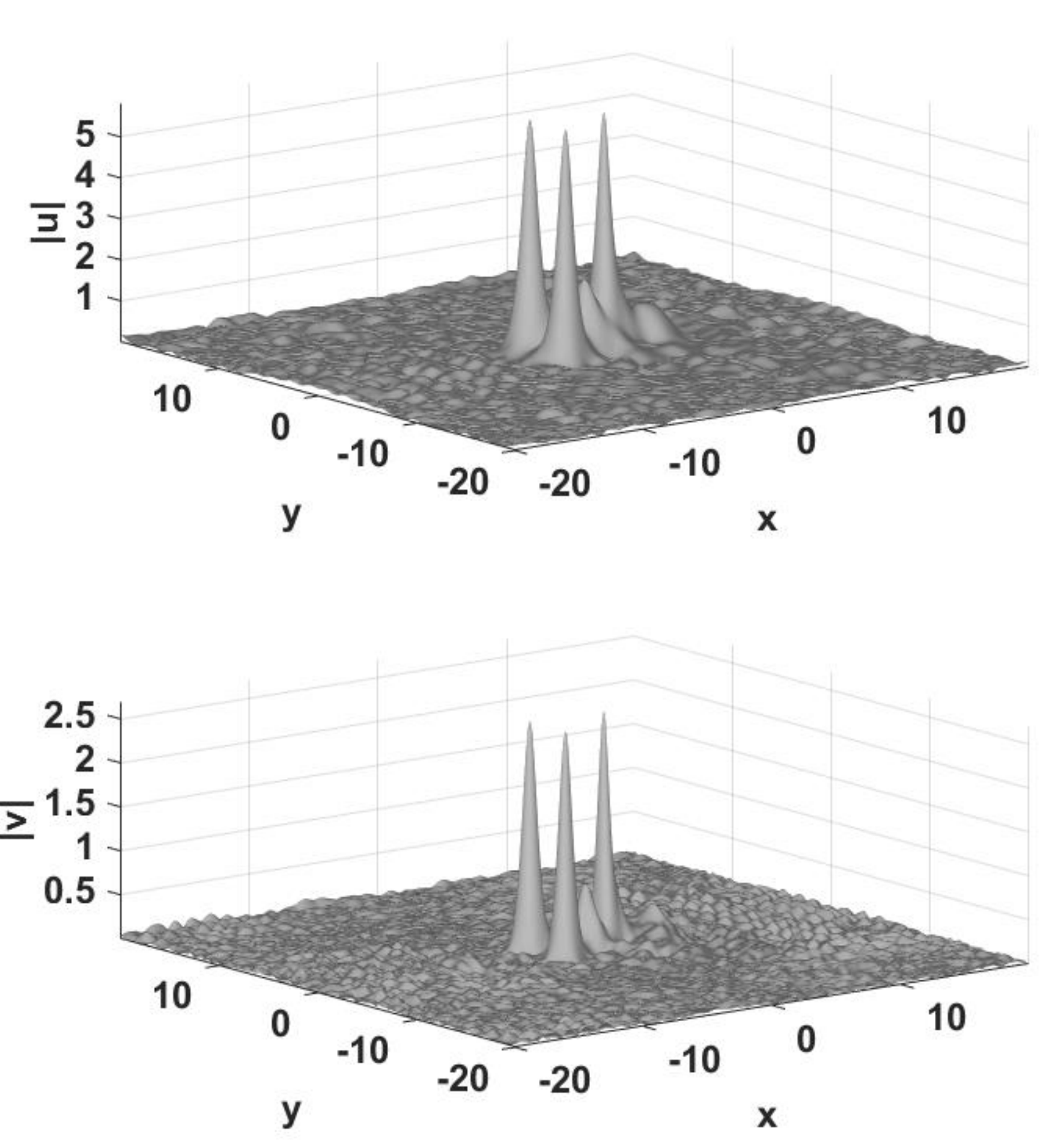}
	} \subfloat[]{
		\includegraphics[width=0.49\textwidth]{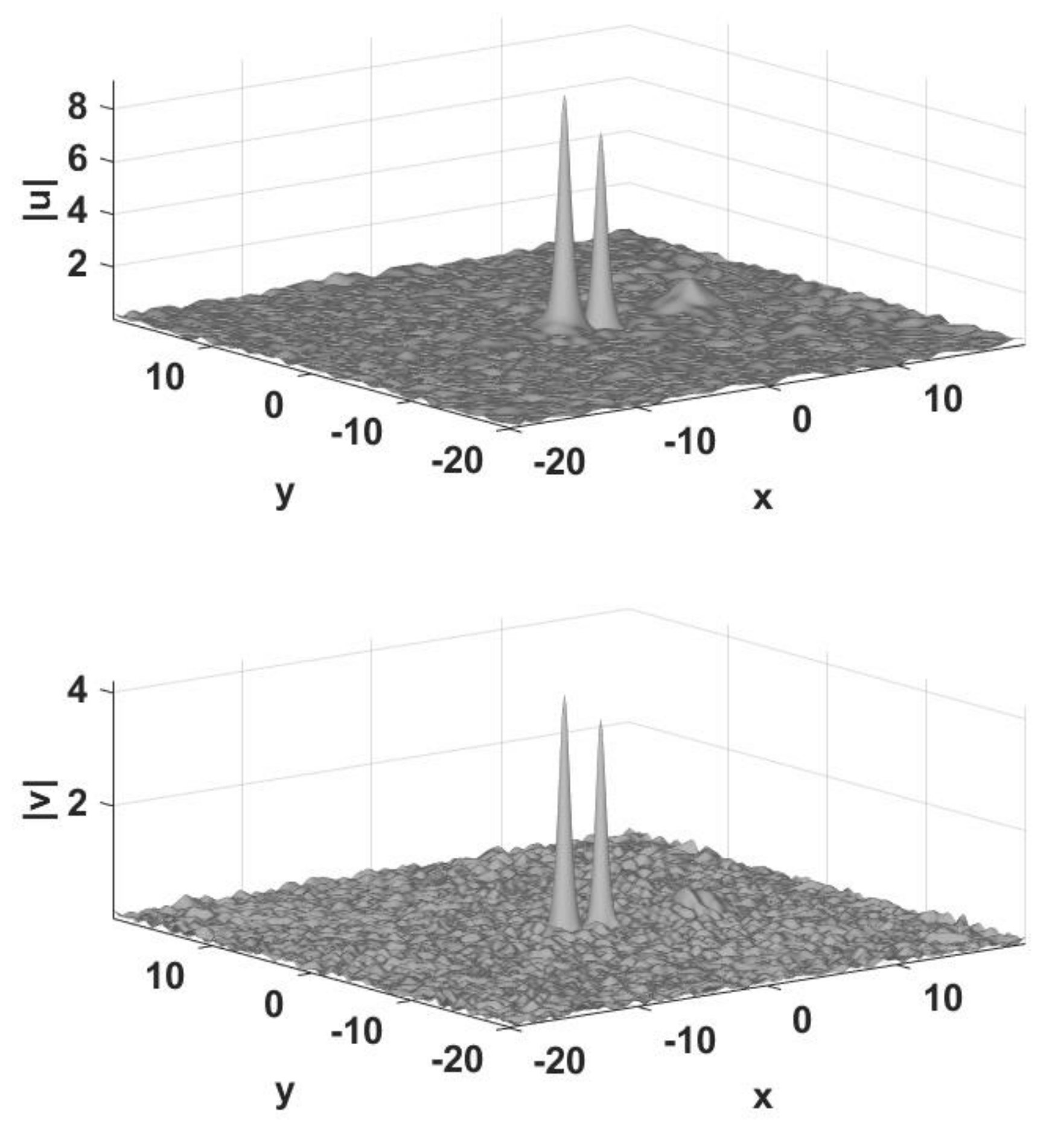}
	}
\caption{Instability-induced spontaneous transformation of a set of seven
solitons into a stable two-soliton state, for parameters $Q=1$, $\protect%
\alpha =0.15$, $R=2.3438$, and the propagation constant of the initial
configuration $k=1.1$. Panels (a), (b), (c) and (d) display, respectively,
shapes of the absolute values of the FF and SH fields at $z=20$, $60$, $80$
and $100$.}
\label{fig:Unstable_2d_7pulses_examples}
\end{figure}

More complex symmetric patterns were found too, including a $10$-soliton set
shown in Fig. (\ref{fig:Unstable_10pulses_symmetric}). However, all the
symmetric sets constructed of more than seven solitons were found to be
unstable. In direct simulations, unstable multi-soliton complexes
spontaneously transform themselves into simpler stable ones -- see, for
instance, the transformation of an unstable seven-soliton set into a stable
two-soliton one in Fig. \ref{fig:Unstable_2d_7pulses_examples}.

\subsection{Asymmetric fundamental modes}

The same three-peak modulation profile may also support soliton sets which
break the triangular symmetry. The simplest among them are asymmetric
stationary single-soliton states, which are shifted from the central
position along the line connecting the center of the three-peak structure
and one of the peaks, in the direction opposite to that towards the peak
(see Fig. \ref{fig:AsymmetricOneandTwoPeaks}(a)). It is possible to predict
the shift analytically, making use of the $\chi ^{(2)}$ term in Hamiltonian $%
H$ (\ref{H}). Indeed, assuming that the soliton has a size much smaller than
its distance $\rho \approx R$ from the singular-modulation peak, the energy
of the soliton's attraction to a particular peak can be written as%
\begin{equation}
E_{\mathrm{attr}}\approx -E_{0}\rho ^{-\alpha },~E_{0}=\int \int \mathrm{Re}%
\left\{ \varphi ^{2}(x,y)\psi (x,y\right\} dxdy,  \label{E}
\end{equation}%
where $\rho $ is the distance of the soliton's center from the peak, and $%
\varphi (x,y)$ and $\psi (x,y)$ are components of stationary solution (\ref%
{stationary}) for the soliton. In fact, it may be taken as the standard
solution for the 2D soliton in the uniform medium \cite%
{Quadratic_solitons_Malomed,Quadratic_solitons}, with quadratic coefficient $%
\chi ^{(2)}=3R^{-\alpha }$, as per Eq. (\ref{R}). The corresponding
attraction force is%
\begin{equation}
F=-\partial E_{\mathrm{attr}}/\partial \rho =-E_{0}\rho ^{-(1+\alpha )}.
\label{F}
\end{equation}

\begin{figure}[tbp]
\centering
{\ \includegraphics[width=0.7\textwidth]{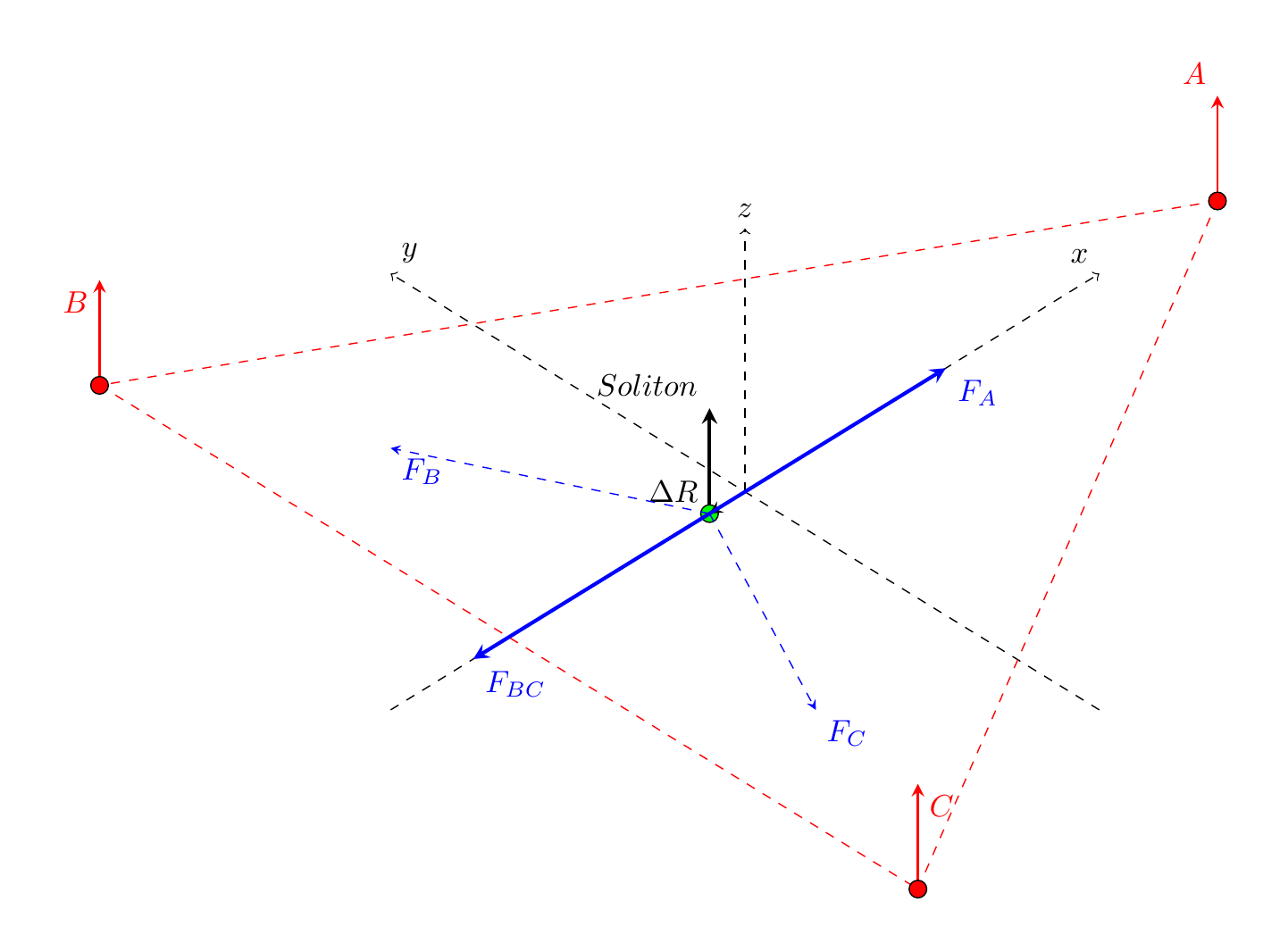} }
\caption{(Color online) The force-equilibrium diagram for the soliton
shifted from the midpoint by distance $\Delta R$, see details in the text.}
\label{fig:ForceEquilibrium}
\end{figure}

Then, a possible equilibrium position of the soliton, treated as a
quasi-particle shifted from the center by distance $\Delta R$, along the
axis connecting the central point and the peak, which corresponds to $\rho
=R+\Delta R$, can be predicted by the condition of the balance between
forces $F_{A}$ and $F_{B}$, $F_{C}$ attracting the soliton to the three
peaks, as shown in Fig. \ref{fig:ForceEquilibrium}. An elementary
consideration gives rise to the balance condition in the following form:%
\begin{equation}
\left( 1+\delta \right) ^{-(1+\alpha )}=(1-2\delta )\left( \frac{3}{4}%
+\left( \frac{1}{2}-\delta \right) ^{2}\right) ^{-\left( 1+\alpha /2\right)
},  \label{balance}
\end{equation}%
where $\delta \equiv (\Delta R)/R$, the right-hand side in Eq. (\ref{balance}%
) being produced by the projection of the forces of the attraction to two
other peaks onto the axis connecting the midpoint and the first peak, see
Fig. \ref{fig:ForceEquilibrium}. Assuming $\alpha \ll 1$ (which
approximately holds for values considered here, such as $\alpha =0.15$), an
expansion of Eq. (\ref{balance}) in powers of small $\alpha $ and presumably
small $\delta $ yields a simple lowest-order result for the asymmetric
equilibrium position of the soliton: $\delta =\alpha /2$. In particular, for
$\alpha =0.15$ this result is $\delta =0.075$, while a numerical solution of
Eq. (\ref{balance}) for the same $\alpha =0.15$ gives $\delta =0.068$. The
full numerical solution for the single asymmetrically placed soliton,
displayed in Fig. \ref{fig:AsymmetricOneandTwoPeaks}(a), yields $\delta _{%
\mathrm{num}}=0.069$, which corroborates the analytical prediction quite
well.

An additional straightforward consideration demonstrates that the so
predicted equilibrium position is a local minimum of the potential
corresponding to forces given by Eq. (\ref{F}), i.e., this position is
stable, which is corroborated by the numerical results. In this connection,
it is relevant to mention that the obvious central equilibrium position, at $%
\delta =0$, which represents the symmetric single-soliton state, corresponds
to a local maximum of the same potential, hence it is formally unstable,
while in the numerical simulations these states are stable (see Fig. \ref%
{fig:FundamentalStabilityMap}(a) below). However, in reality all stable
symmetric single-soliton modes exist with small values of $k$ in Fig. \ref%
{fig:FundamentalStabilityMap}(a), hence they are quite broad, covering
nearly the entire triangle formed by the three singular-modulation peaks, as
clearly seen in Fig. \ref{fig:Stable_1pulse_symmetric}; for this reason, the
above treatment of the soliton as a quasi-particle is not relevant in this
case. On the other hand, values of $k$ for stable asymmetric single-soliton
states in Fig. \ref{fig:FundamentalStabilityMap}(a) are larger than for
their symmetric counterparts, making these solitons more compact objects (as
corroborated by Fig. \ref{fig:AsymmetricOneandTwoPeaks}(a)), for which the
quasi-particle approximation is appropriate.

Stable two-soliton asymmetric sets were found too, for relatively large
values of $R$, with two singularity peaks carrying solitons pinned to them,
while the third one remains nearly \textquotedblleft empty", see Fig. \ref%
{fig:AsymmetricOneandTwoPeaks}(b). The largest number of individual solitons
in stable asymmetric patterns is three (unlike seven for the symmetric
sets). Larger asymmetric complexes were found too, such as a slightly bent
five-soliton string shown in Fig. (\ref{fig:Unstable_5pulses_asymmetric}),
but they all are unstable. Note that, similar to the symmetric multi-soliton
sets, adjacent solitons in the string have opposite signs in their FF
components.

\begin{figure}[tbp]
\centering\subfloat[]{
		\includegraphics[width=0.45\textwidth]{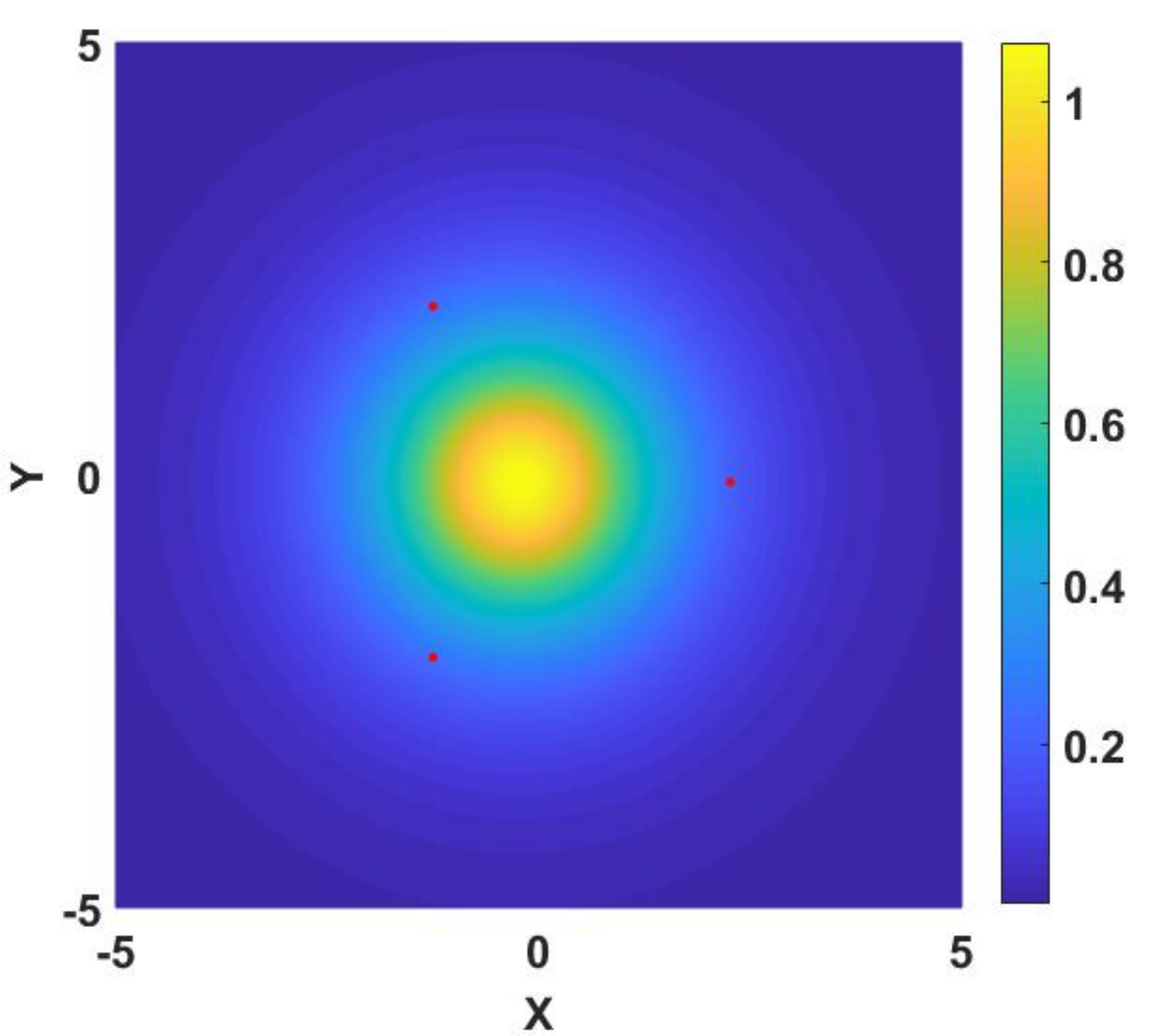}
	} \subfloat[]{
	    \includegraphics[width=0.45\textwidth]{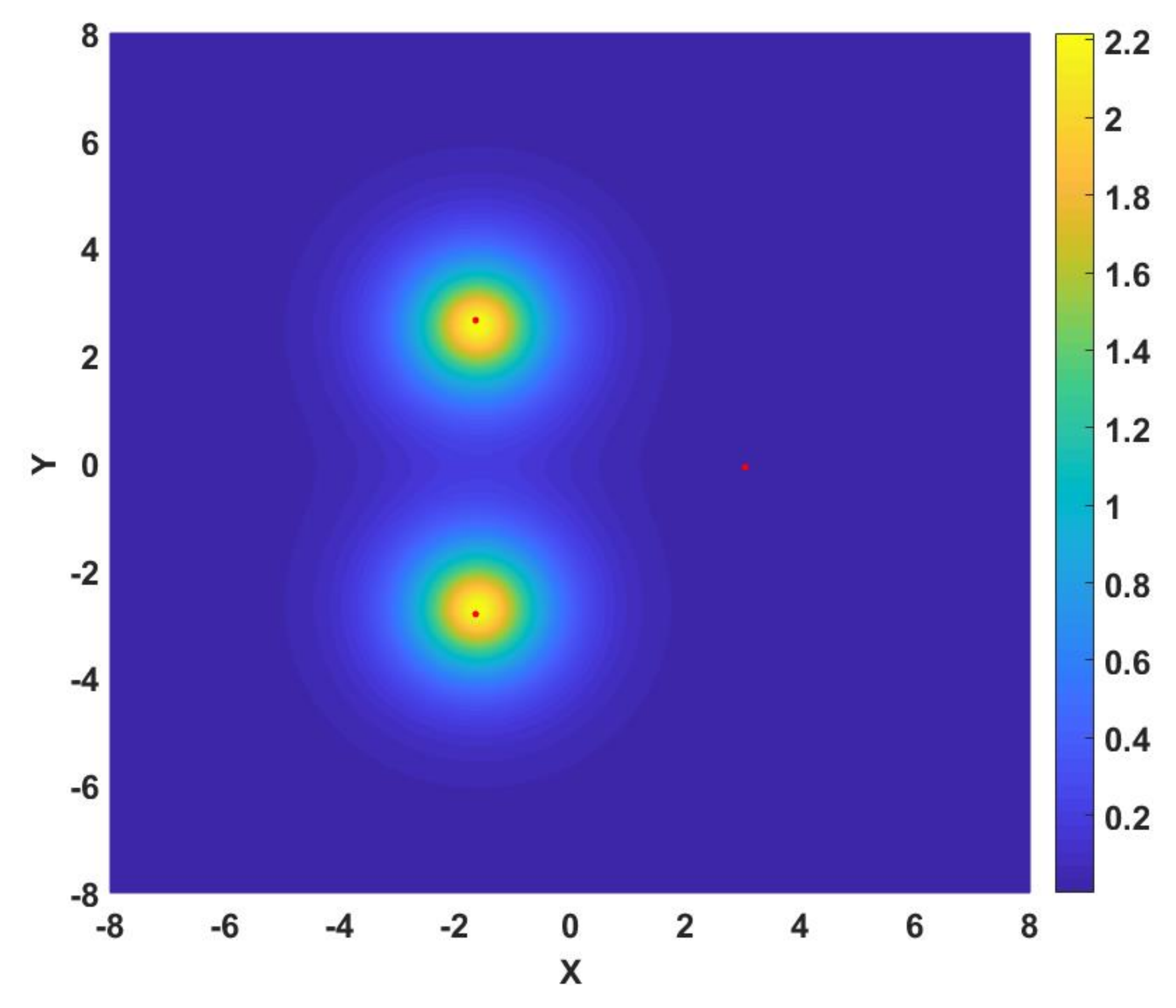}
	}
\caption{(Color online) (a) The FH field of \ a stable asymmetric
single-soliton state, whose stationary position is shifted off the center.
The parameters are $\protect\alpha =0.15$, $Q=1$, and $R=2.3438$, the
respective propagation constant being $k=0.35$ (b) The same for a stable
asymmetric two-soliton set, with $\protect\alpha =0.15$, $Q=1$, $R=3.1250$,
and $k=1.1$.}
\label{fig:AsymmetricOneandTwoPeaks}
\end{figure}

\begin{figure}[tbp]
\centering
\subfloat[]{
		\includegraphics[width=0.45\textwidth]{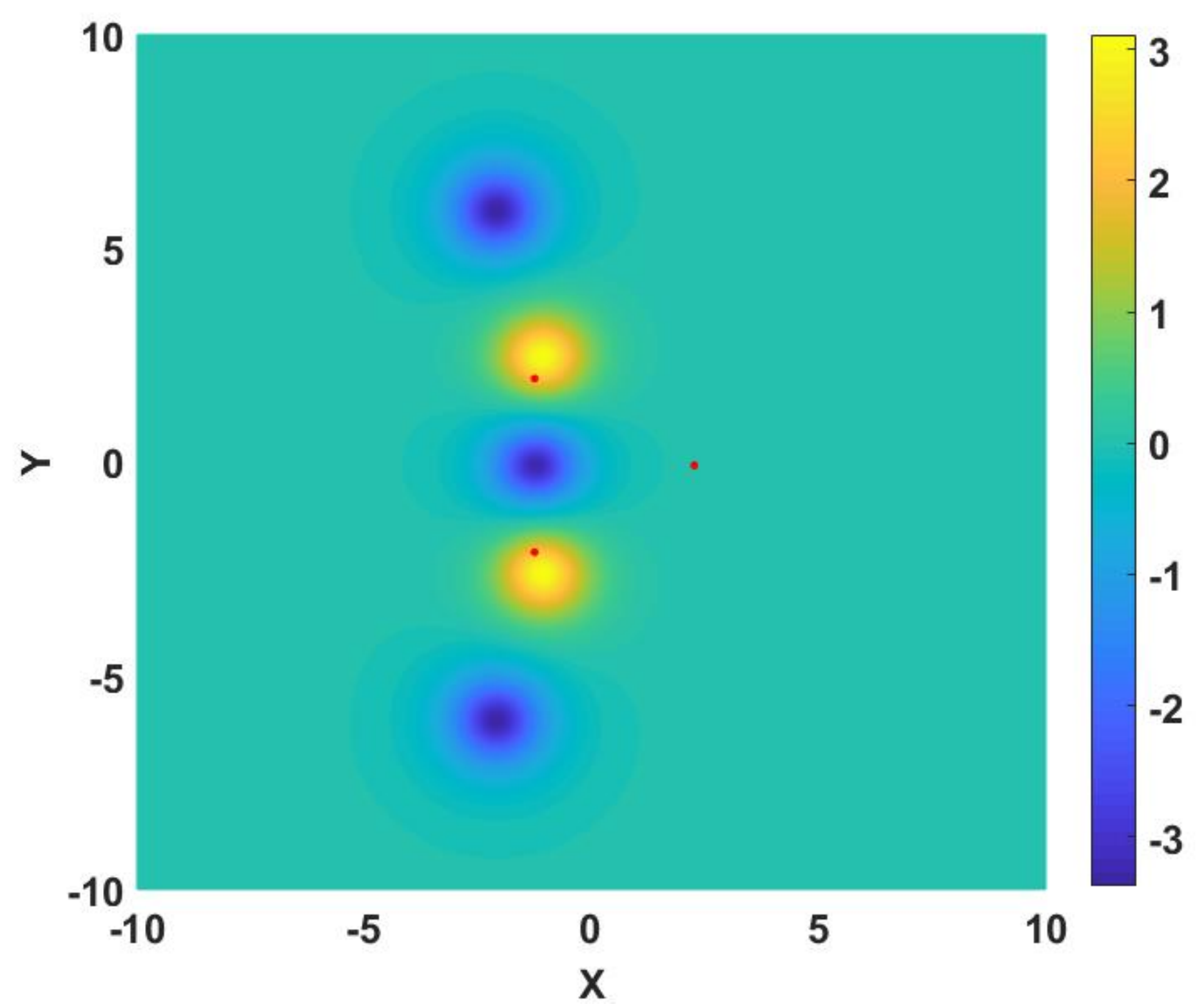}
	} \subfloat[]{
		\includegraphics[width=0.45\textwidth]{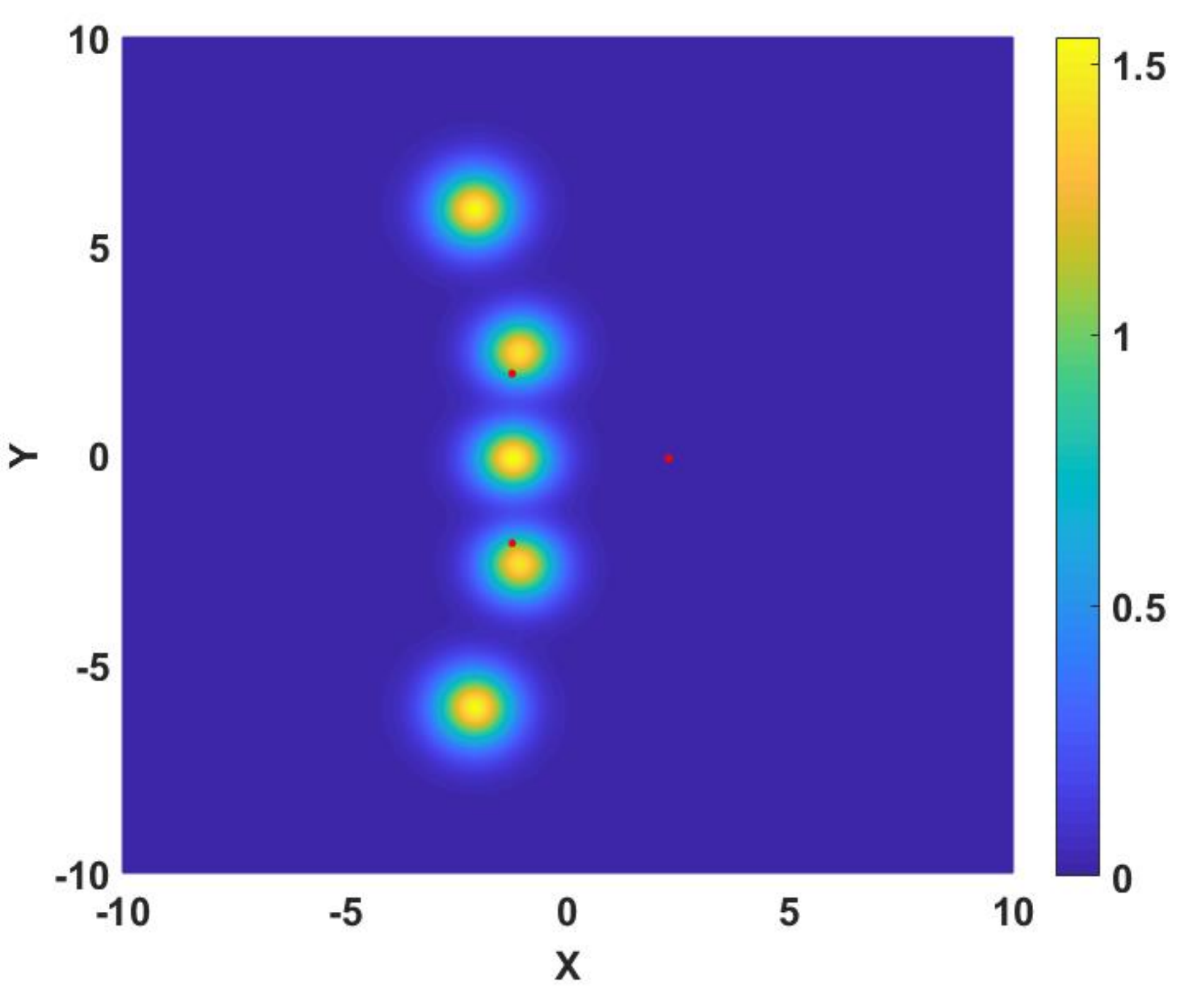}
	}
\caption{(Color online) An unstable asymmetric set of five solitons, for
parameters $\protect\alpha =0.15$, $Q=1$,and $R=2.3438$, the propagation
constant being $k=1.1$. Panels (a) and (b) display the FH and SH fields, $%
\protect\varphi \left( x,y\right) $ and $\protect\psi \left( x,y\right) $,
respectively.}
\label{fig:Unstable_5pulses_asymmetric}
\end{figure}

Results obtained for the existence and stability of various single- and
multi-soliton complexes are summarized in Fig. (\ref%
{fig:FundamentalStabilityMap}), in the form of $P(k)$ dependences for
different species of the multisoliton sets, separately for symmetric and
asymmetric ones (panels (a) and (b), respectively; recall $P$ is the total
power defined by Eq. (\ref{P})). The results are collected in this figure
for $\alpha =0.15$, $Q=1$, and $R=2.3438$, which adequately represents a
generic case. A general trend is destabilization with the increase of the
total power, but there are exceptions, such as families of symmetric sets of
four and seven solitons in panel (a). Another general trend is the
destabilization with the increase of the number of solitons in the sets: as
mentioned above, symmetric and asymmetric complexes cannot be built of more
than seven and three individual solitons, respectively.
\begin{figure}[tbp]
\centering
\subfloat[]{
		\includegraphics[width=0.9\textwidth]{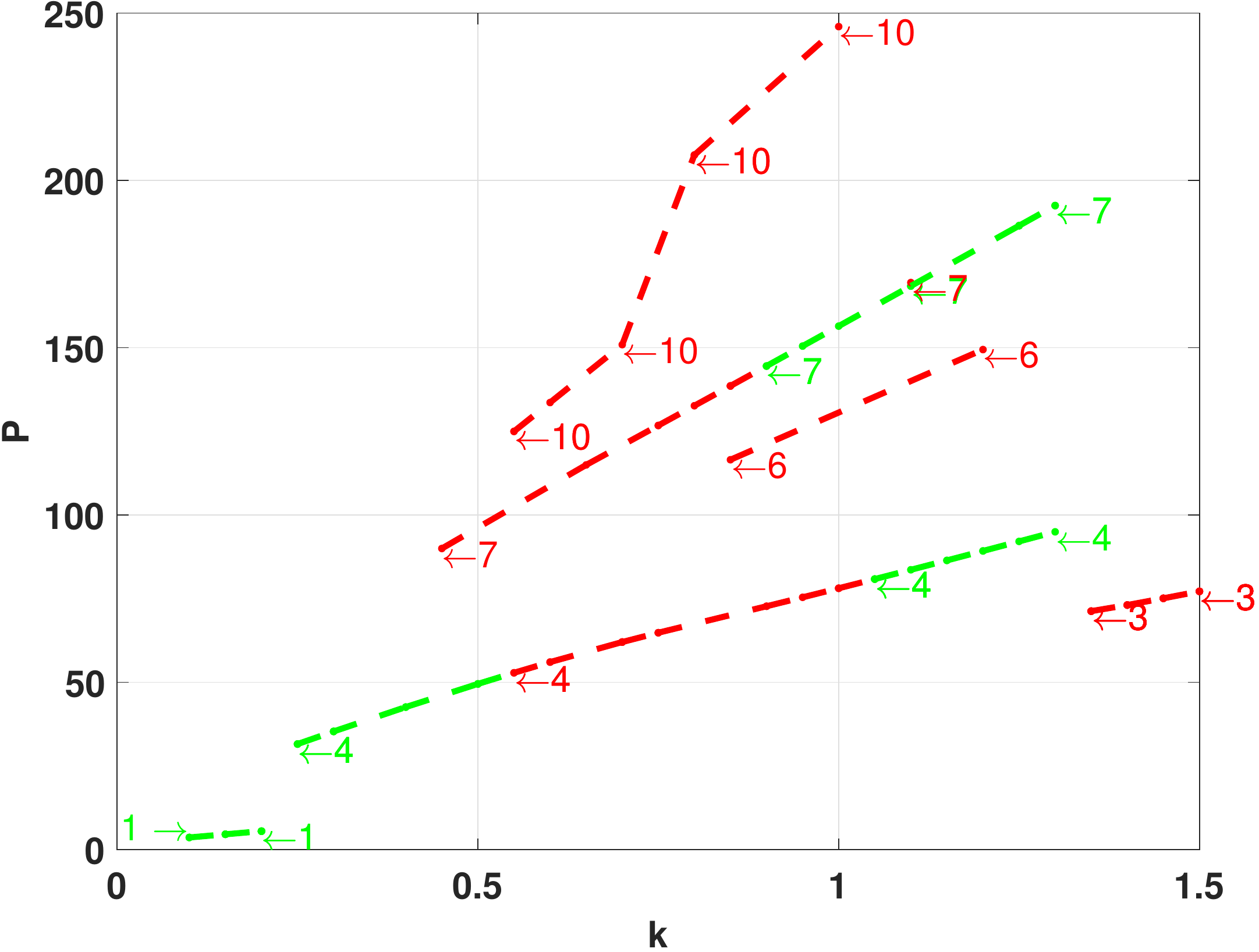}
	} \newline
\subfloat[]{
		\includegraphics[width=0.9\textwidth]{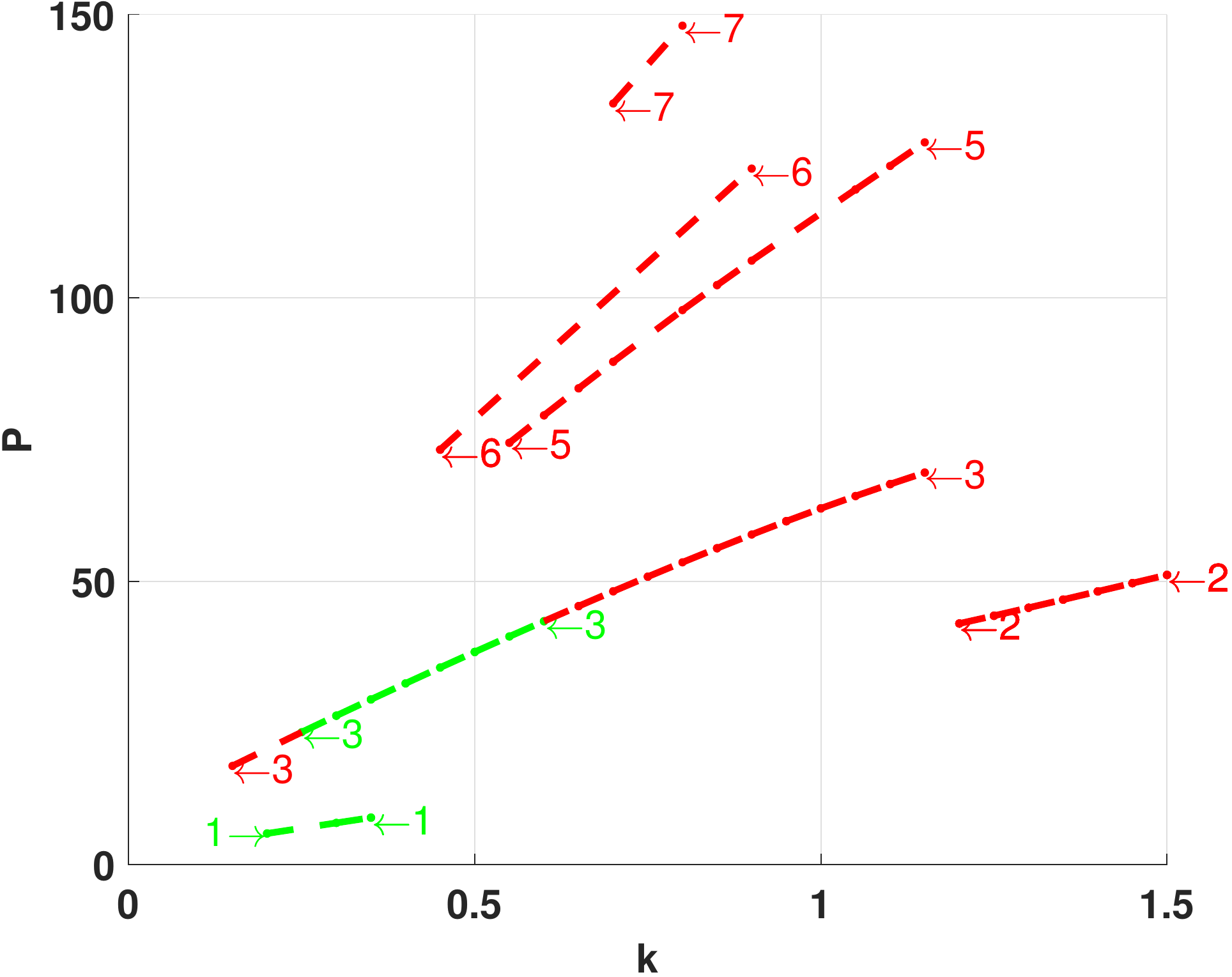}
	}
\caption{(Color online) Typical existence and stability maps for single- and
multi-solitons states, shown in the plane of the propagation consonant, $k$,
and total power ($P$, see Eq. (\protect\ref{P})), for $\protect\alpha %
=0.15,Q=1$ and $R=2.3438$. Numbers attached to edges of the $P(k)$ lines,
and to their internal breakup points (see the top line in (a) labeled
\textquotedblleft 10") denote the number of individual solitons in the
respective sets. Green and red colors of the lines and attached numbers
designate stable and unstable sets, respectively. Panels (a) and (b)
separately display the results for symmetric and asymmetric soliton
complexes.}
\label{fig:FundamentalStabilityMap}
\end{figure}

It is worthy to note that the comparison of panels (a) and (b) in Fig. \ref%
{fig:FundamentalStabilityMap} demonstrates bistability in the present
system, which we understand as coexistence of different stable patterns at
equal values of their total powers. First, the interval of values of the
total power in which the asymmetric single-soliton states are stable
overlaps with the stability interval of the symmetric single-soliton modes.
Second, the stability intervals for the asymmetric three-soliton sets and
symmetric four-soliton ones nearly coincide too.

Lastly, additional numerical results, obtained by varying distance $R$ of
each peak from the center, demonstrate that all two and three-soliton sets,
with each soliton pinned to a singularity peak, tend to become stable with
the increase of $R$, for a simple reason that weak interactions between far
separated solitons cannot conspicuously destabilize their bound states. In
particular, the diagram for the asymmetric sets, displayed in Fig. \ref%
{fig:FundamentalStabilityMap}(b), does not feature any stable two-soliton
complexes at $R=2.3438$, while at $R=3.1250$ a stable two-soliton state is
readily found, as shown in Fig. \ref{fig:AsymmetricOneandTwoPeaks}(b).

\section{Vortex-soliton rings}

\begin{figure}[tbp]
\centering
\par
\subfloat[]{
		\includegraphics[width=0.45\textwidth]{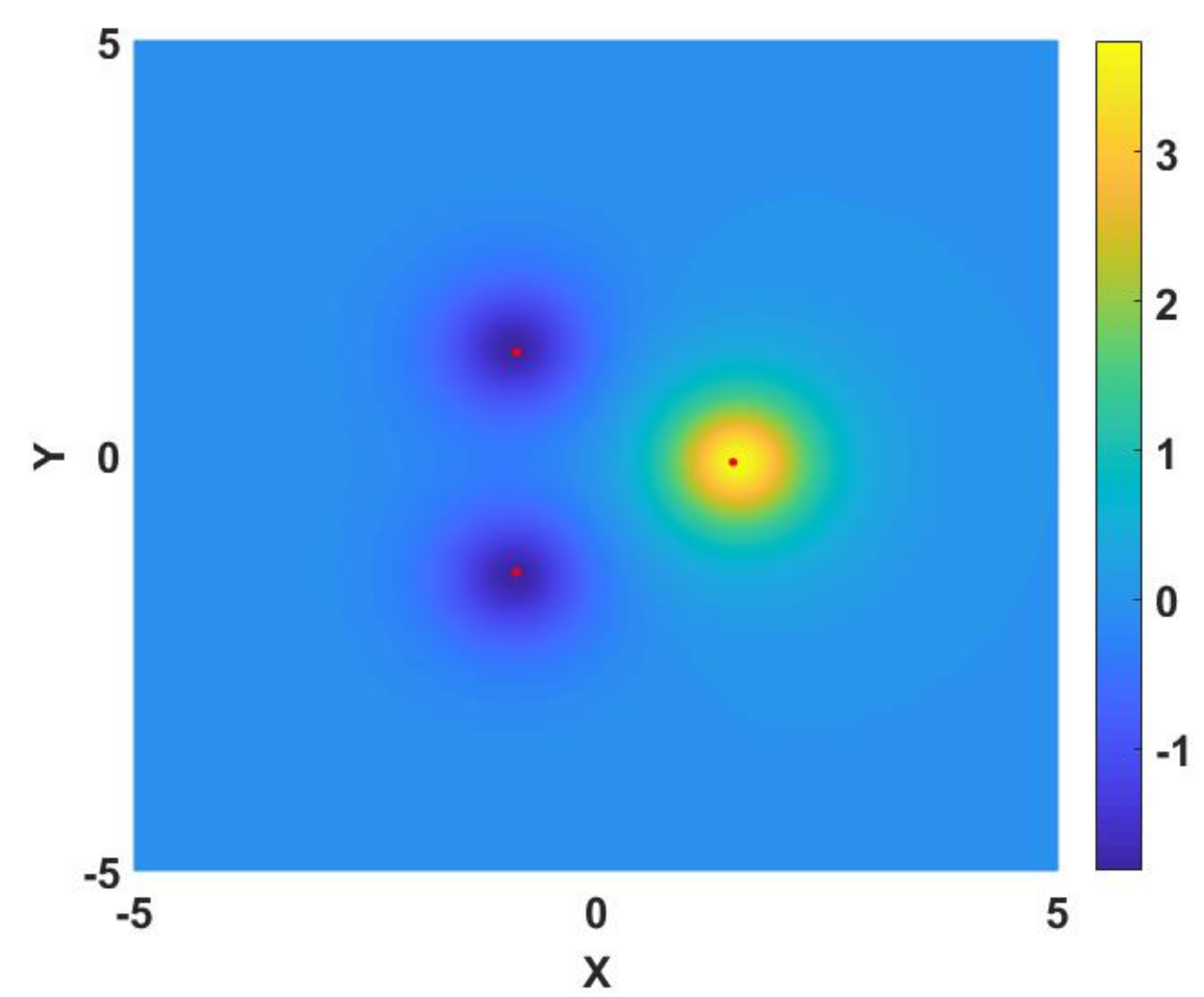}
	} \subfloat[]{
		\includegraphics[width=0.45\textwidth]{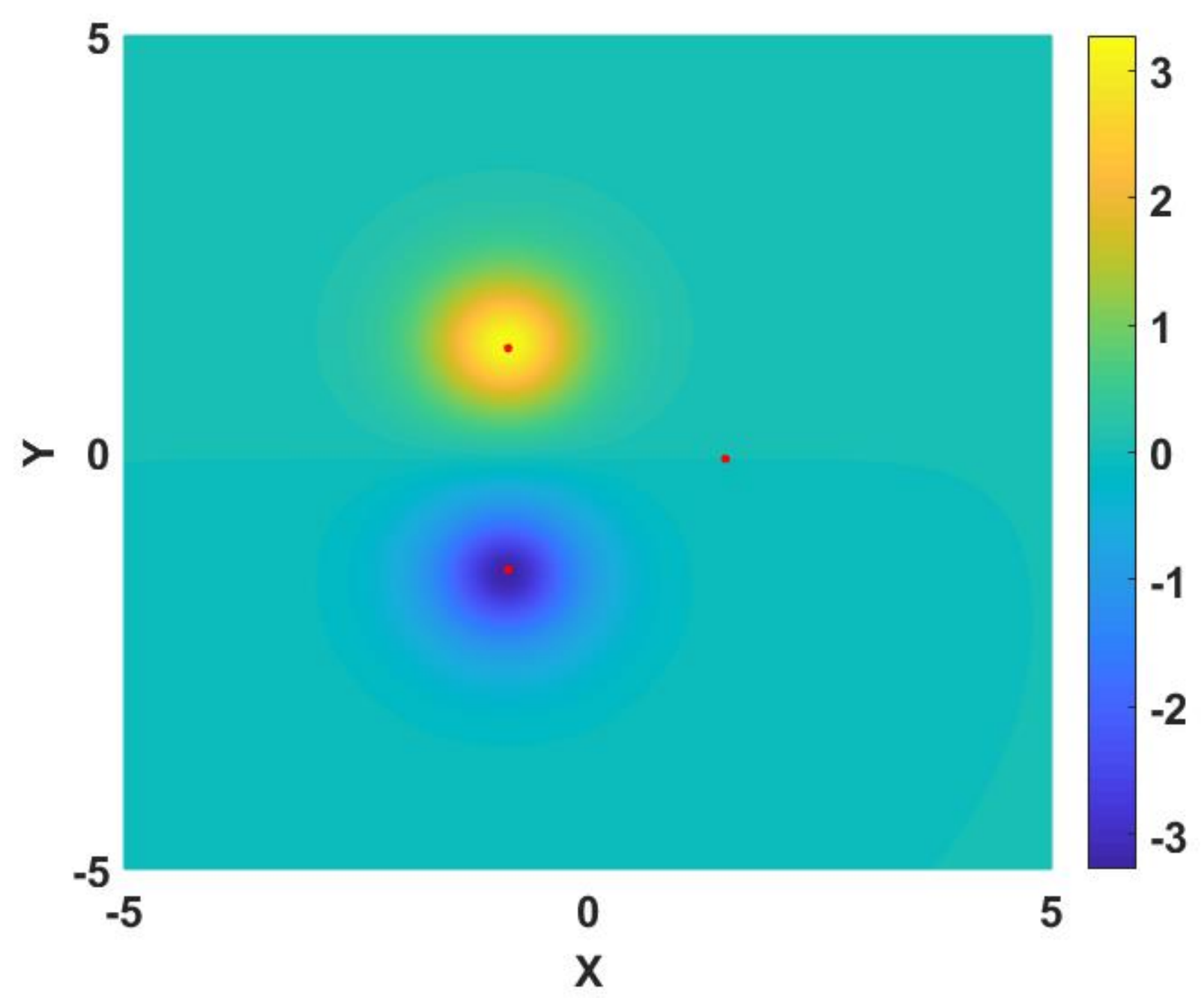}
	}\newline
\par
\subfloat[]{
	\includegraphics[width=0.45\textwidth]{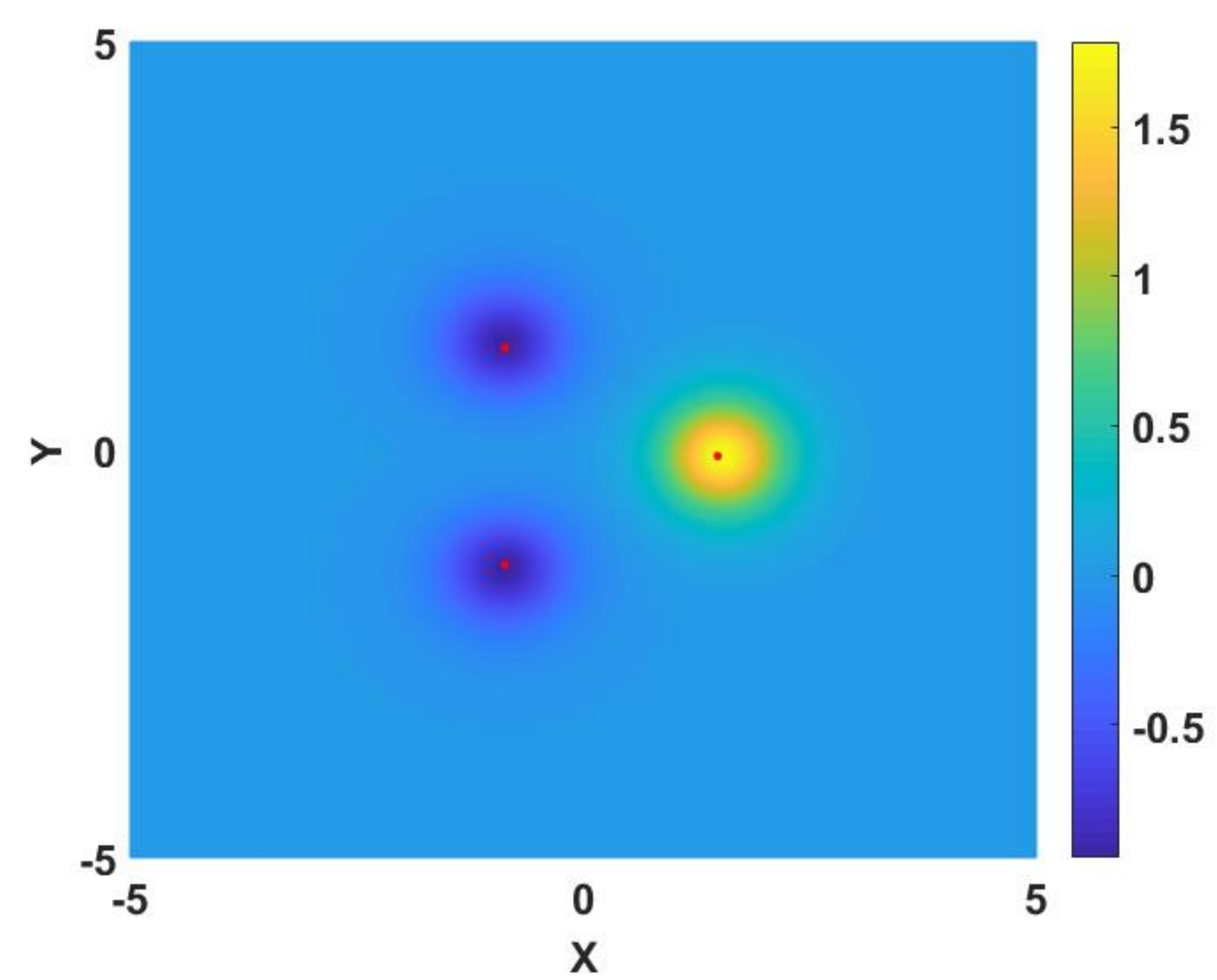}
	} \subfloat[]{
	\includegraphics[width=0.45\textwidth]{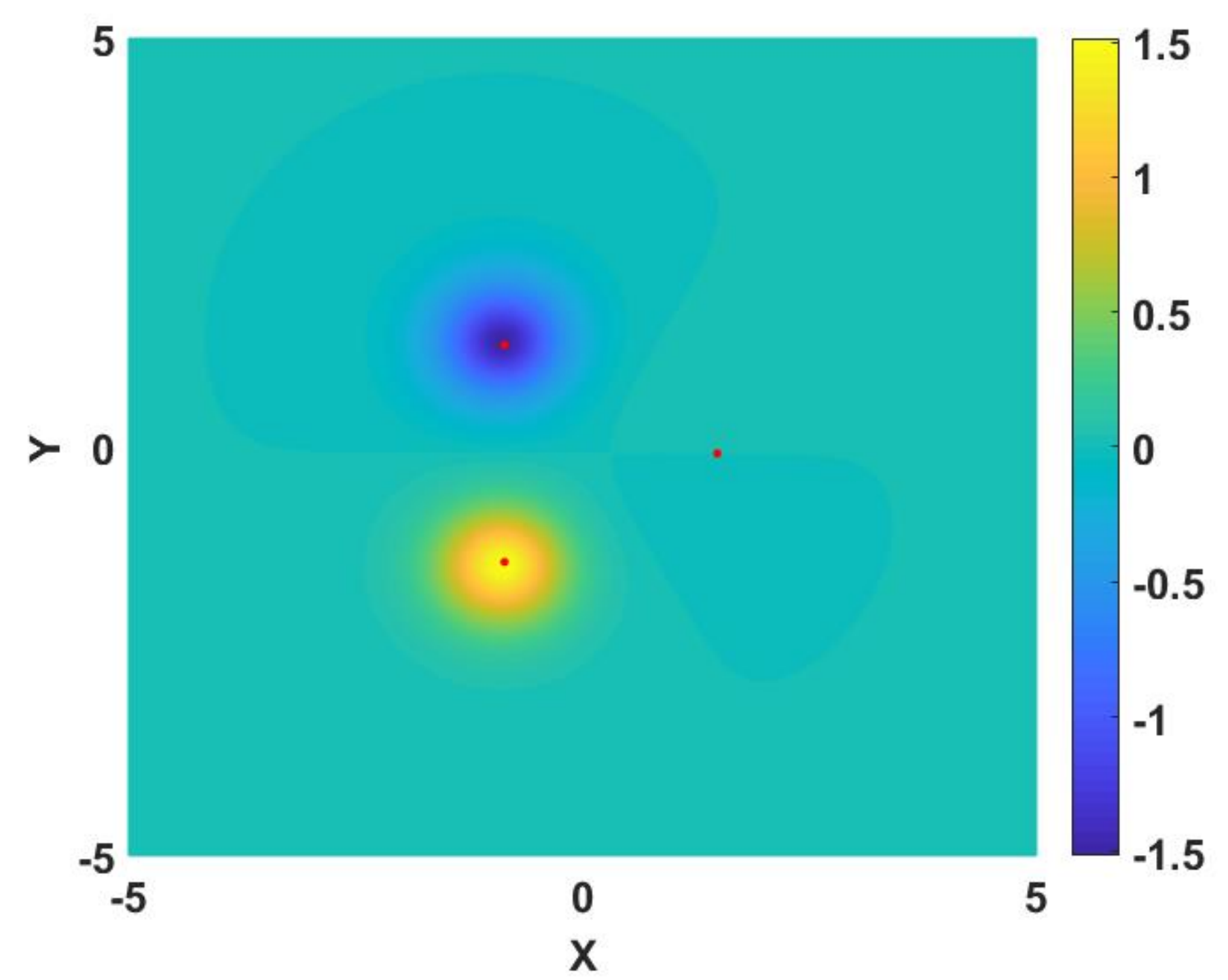}
	}
\caption{(Color online) A stable three-soliton ring with imprinted vorticity
$S=1$, for parameters $\protect\alpha =0.15,Q=1$, $R=1.5625$, and $k=1.5$.
Panels (a) and (b) display the real and imaginary parts of the FH field; (c)
and (d): the same for the SH field.}
\label{fig:Stable_3pulses_vortex_symmetric}
\end{figure}

Three-soliton sets carrying intrinsic vorticity were generated starting from
the input built as a ring-shaped chain of three fundamental solitons
attached to the singular-modulation peaks (A, B, C in Fig. \ref{fig1}), with
phase shift $2\pi /3$ between FF components of adjacent solitons, which
corresponds to the overall phase circulation of $2\pi $ in the FF field (and
$4\pi$ in its SH counterpart), i.e., vorticity $S=1$ \textit{imprinted} onto
the soliton ring:

\begin{gather}
\varphi \left( x,y\right) =\varphi _{A}(x,y)+\varphi _{B}(x,y)e^{2\pi
i/3}+\varphi _{C}(x,y)e^{4\pi i/3},  \notag \\
\psi \left( x,y\right) =\psi _{A}(x,y)+\psi _{B}(x,y)e^{4\pi i/3}+\psi
_{C}(x,y)e^{8\pi i/3},  \label{vort}
\end{gather}%
where subscripts A, B, and C refer to standard fundamental-soliton solutions
with their centers placed at the respective singular peaks (in the last term
of the expression for $\psi$, $e^{8\pi i/3}$ is identical to $e^{2\pi i/3}$%
). An example of the so constructed stable vortex complex is displayed in
Fig. \ref{fig:Stable_3pulses_vortex_symmetric}.

\begin{figure}[tbp]
\centering
{\ \includegraphics[width=0.45\textwidth]{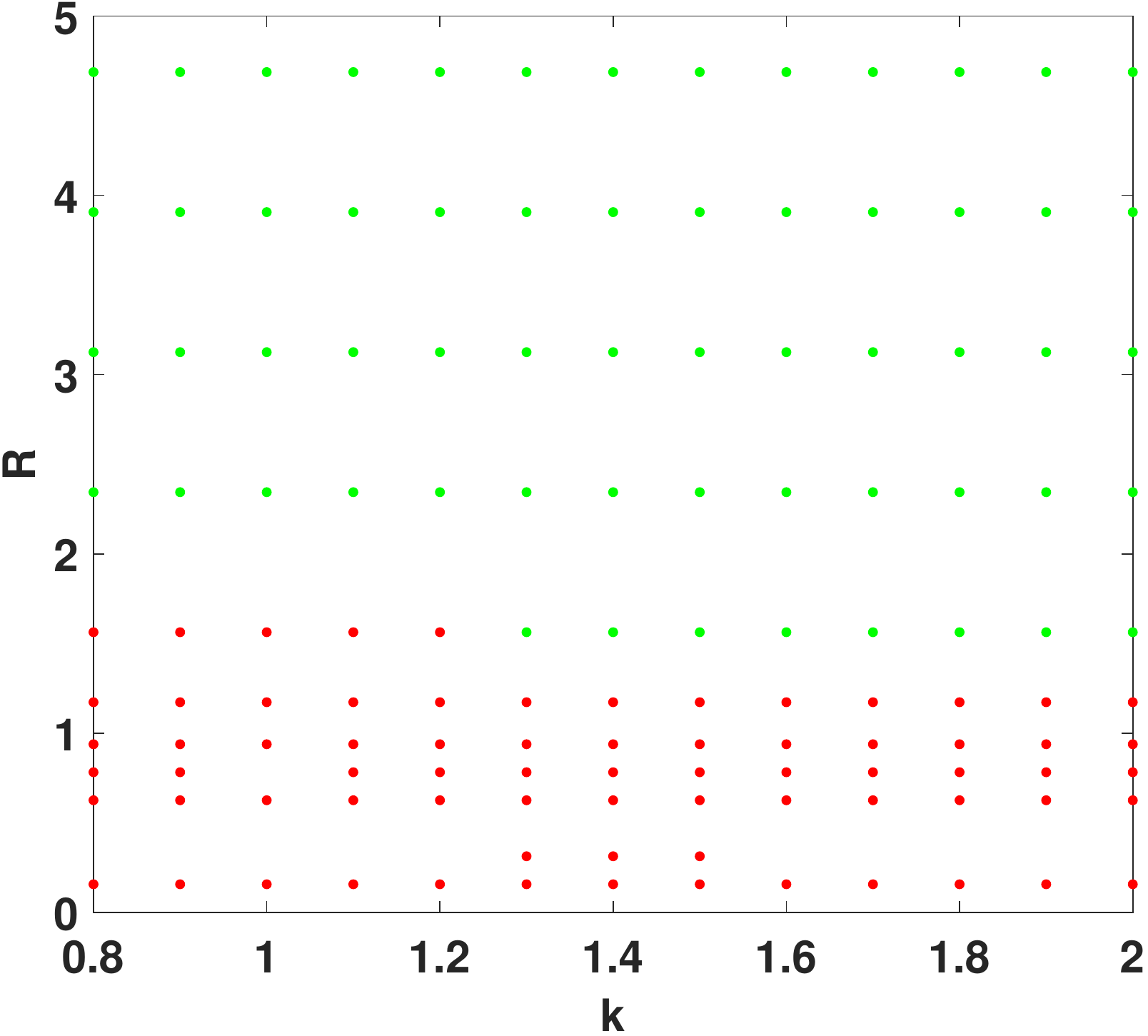} }
\caption{(Color online) The stability diagram for vortex-ring solitons at $%
Q=1$ and $\protect\alpha =0.15$ in the plane of the propagation constant $k$
and the distance of each singularity peak from the center, $R$. Green and
red dots denote, respectively, stable and unstable vortices.}
\label{fig:VortexStabilityRvsK}
\end{figure}

As shown in Fig. \ref{fig:VortexStabilityRvsK}, the stability of vortex
rings essentially depends on size $R$ of the three-peak configuration: as
well as in case of zero-vorticity multisoliton sets (see above), the vortex
rings tend to stabilize themselves with the increase of $R$, which is
explained by weaker interaction between far separates solitons. The
stability region in Fig. \ref{fig:VortexStabilityRvsK} somewhat expands with
the increase of $k$, because this entails shrinkage of each individual
soliton, making its relative size with respect to $R$ smaller, thus also
effectively weakening the interaction between solitons.

Lastly, in direct simulations the vortex rings, which are predicted to be
unstable in terms of eigenvalues of small perturbations, demonstrate
spontaneous transformation into a single stable soliton, with a small
\textquotedblleft remnant" of another one, as shown in Fig. (\ref%
{fig:Unstable_2d_3pulses_vortex_examples})

\begin{figure}[tbp]
\centering
\subfloat[]{
		\includegraphics[width=0.49\textwidth]{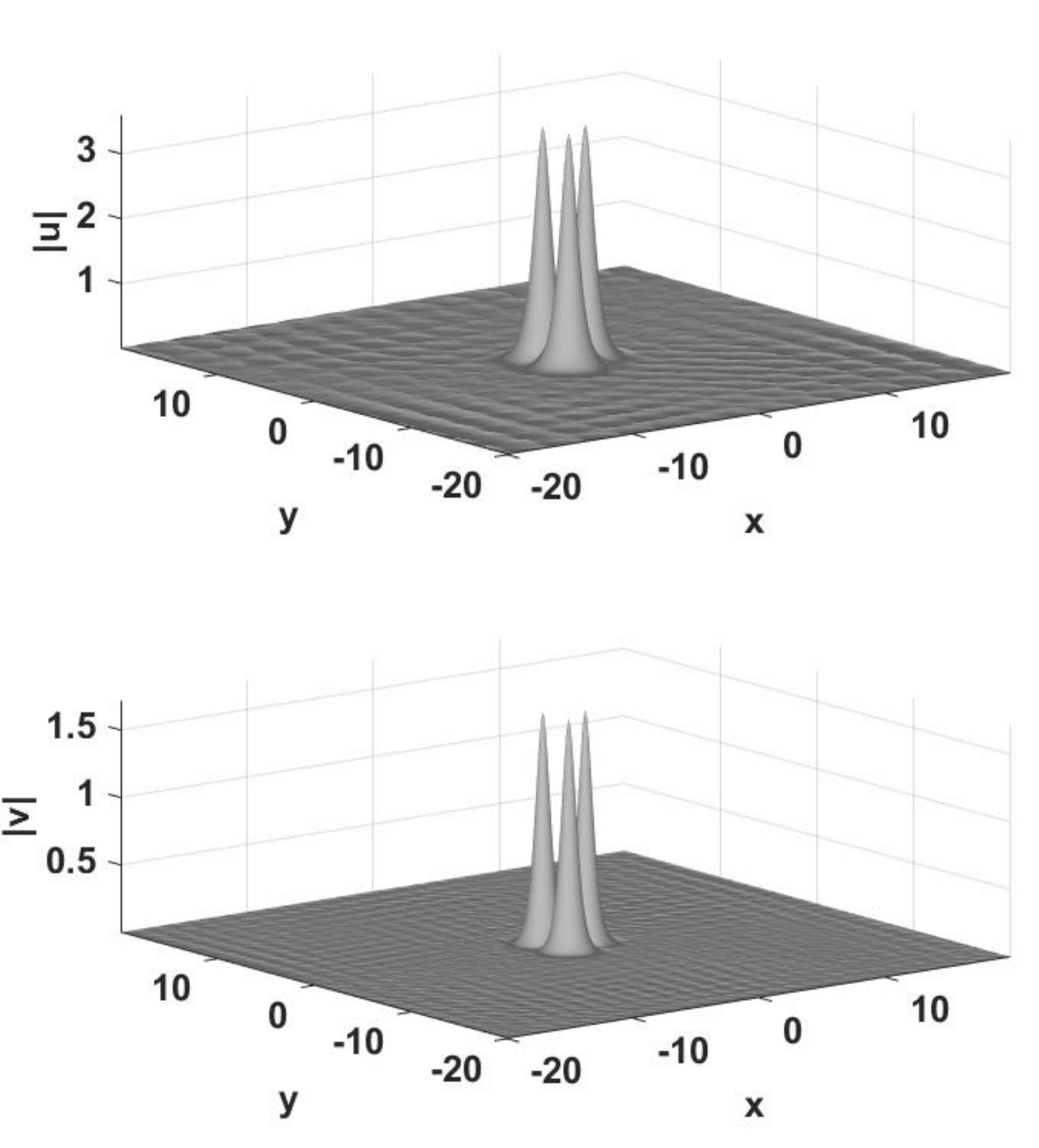}
	} \subfloat[]{
		\includegraphics[width=0.49\textwidth]{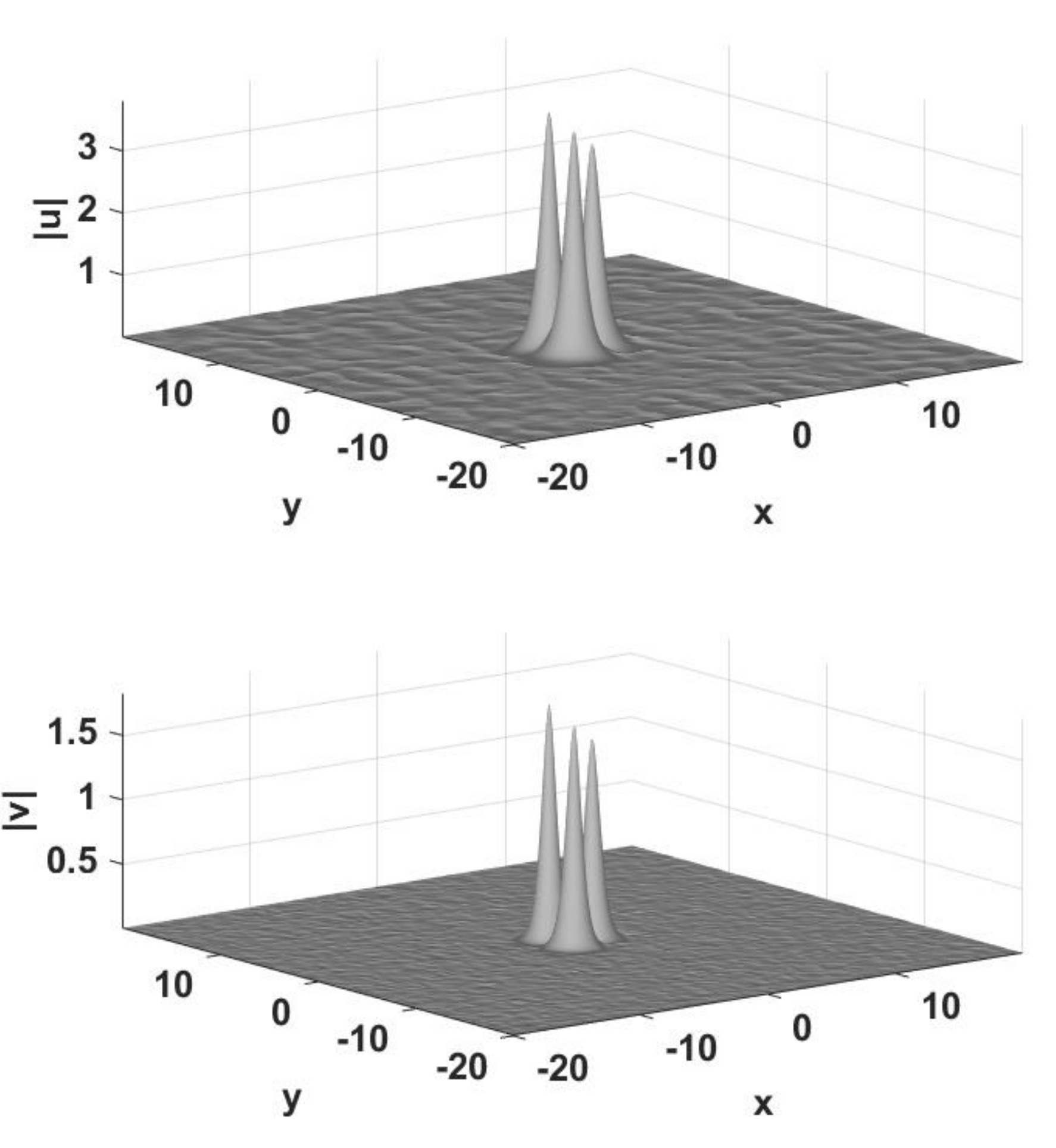}
	} \newline
\subfloat[]{
		\includegraphics[width=0.49\textwidth]{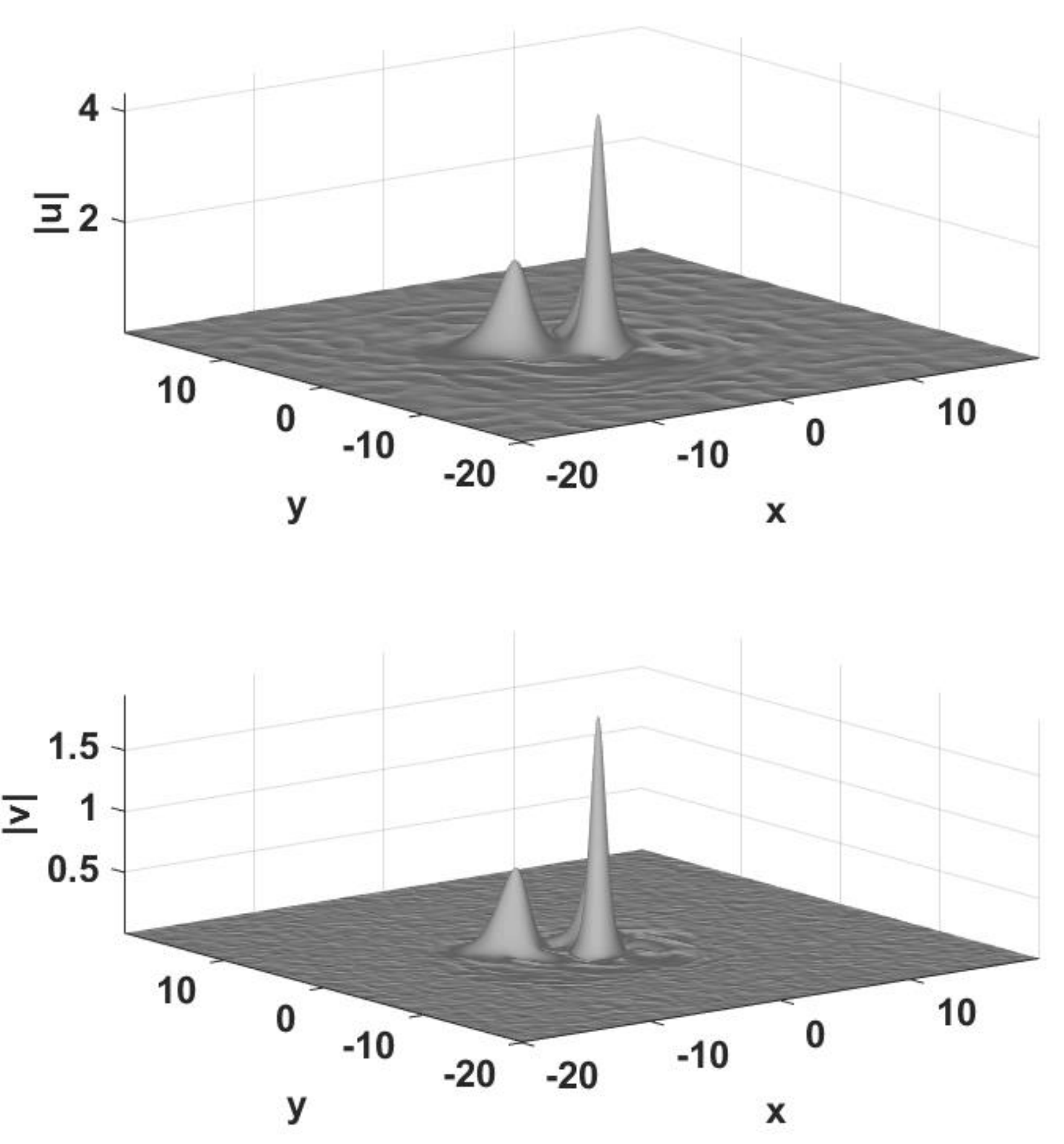}
	} \subfloat[]{
		\includegraphics[width=0.49\textwidth]{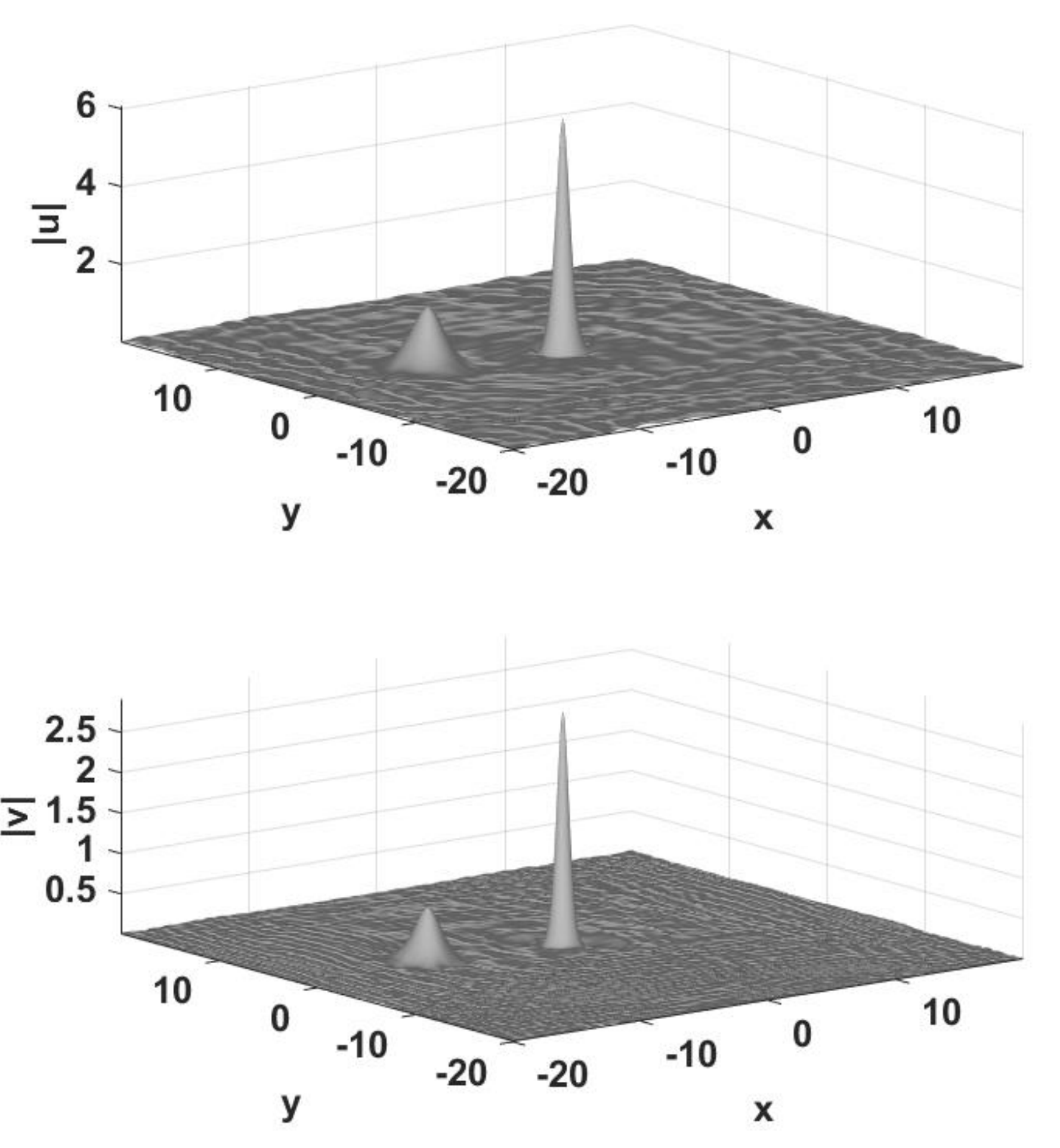}
	}
\caption{Unstable evolution of a vortex ring, originally built of three
individual solitons, with $k=1.1$. Other parameters are $Q=1$, $\protect%
\alpha =0.15$, and $R=1.5625$. Panels (a), (b), (c) and (d) display,
respectively, the absolute value of the FF and SH fields at $z=20$, $60$, $%
80 $ and $100$.}
\label{fig:Unstable_2d_3pulses_vortex_examples}
\end{figure}

\section{Conclusion}

Using the recent result which demonstrates the existence of stable 2D
solitons in the medium with the quadratic interaction supported by the local
$\chi ^{(2)}$ coefficient which is subject to the locally singular
modulation, we have introduced the system with the equilateral triangle of
three singular-modulation peaks. This system makes it possible to predict
new 2D multi-soliton patterns, including ones symmetric and asymmetric with
respect to the underlying triangular structure, and introduce vortex-ring
patterns. The stability of the stationary solutions was identified through
the computation of the respective perturbation eigenvalues, and verified by
direct simulations. The asymmetric position of the single-soliton state,
shifted off the system's center, was accurately predicted analytically.
Zero-vorticity (fundamental) multi-soliton sets were built with the number
of solitons from $1$ to $10$. Symmetric sets constructed of $1$, $4$, and $7$
solitons have their stability regions, while stable asymmetric patterns may
contain $1$, $2$, or $3$ solitons. Symmetric and asymmetric sets composed of
$1$ or $3$ and $4$ solitons feature mutual bistability. All sets elongated
in a particular direction feature alternating signs of the
fundamental-frequency components of constituent solitons in that direction.
Unstable patterns spontaneously transform themselves into stable ones, with
a smaller number of individual solitons in them. Vortex rings, composed of
three solitons, are stable if interactions between individual solitons are
sufficiently weak.

These results present a contribution to the vast topic of pattern formation
in spatial-domain nonlinear optics \cite{NNR}-\cite{Cid}. As an extension of this
work, it may be interesting to consider a system of three waves coupled by
the singularly modulated quadratic Type-II interaction.

\section*{Acknowledgments}

We thank Prof. C. L. P. Lambruschini for the invitation to submit a contribution
to this Special Issue of EPJ - ST. This work was supported, in a part, by the 
Israel Science Foundation through grant No. 1287/17.

\section*{Authors' contributions}
The work was designed by B.A.M., who was also responsible for the analytical part.
V.L. has performed the numerical computations. Both authors equally contributed to
drafting the paper.

\end{document}